\documentclass[sigconf]{acmart}
\pdfoutput=1

\usepackage{paper}

\newcommand{\tudparagraph}[1]{\leavevmode\unskip\\[-.8em]\noindent\textbf{#1.}}
\newcommand{\RotText}[1]{\rotatebox{90}{\parbox{1cm}{#1}}}

\newcommand{\cC}{\ensuremath{\mathcal{C}}\xspace}

\newcommand{\cP}{\ensuremath{\mathcal{P}}\xspace}

\newcommand{\cT}{\ensuremath{\mathcal{T}}\xspace}

\makeatletter
\newcommand{\oset}[3][-.1ex]{%
  \mathrel{\mathop{#3}\limits^{
    \vbox to#1{\kern-2\ex@
    \hbox{$\scriptstyle#2$}\vss}}}}
\makeatother

\makeatletter
\newcommand{\cset}[3][-.1ex]{%
  \mathrel{\mathop{#3}\limits^{
    \hspace{-.5ex}\vbox to#1{\kern-2\ex@
    \hbox{$\scriptstyle#2$}\vss}}}}
\makeatother

\newcommand{\en}[1]{\enskip #1 \enskip}

\newcommand{\ra}{\ensuremath{\rightarrow}\xspace}
\newcommand{\pa}{\ensuremath{\rightharpoonup}\xspace}

\newcommand{\Feat}{\ensuremath{F}\xspace}
\newcommand{\ValidFeat}{\ensuremath{\mathcal{V}}\xspace}
\newcommand{\Effects}{\ensuremath{\mathcal{E}}\xspace}
\newcommand{\CEffects}{\ValidFeat{\setminus}\Effects}
\newcommand{\plainCauses}{\ensuremath{\mathsf{Causes}}\xspace}
\newcommand{\Causes}{\ensuremath{\plainCauses(\Effects,\ValidFeat)}\xspace}
\newcommand{\CCauses}{\ensuremath{\plainCauses(\CEffects,\ValidFeat)}\xspace}

\newcommand{\mCauses}{\ensuremath{m\Causes}\xspace}
\newcommand{\mCCauses}{\ensuremath{m\CCauses}\xspace}

\newcommand{\true}{\mathtt{true}}
\newcommand{\false}{\mathtt{false}}
\newcommand{\Sem}[1]{\llbracket#1\rrbracket}

\newcommand{\Real}{\mathbb{R}}
\newcommand{\power}[1]{{\wp}(#1)}
\newcommand{\supp}{\mathsf{supp}}
\newcommand{\gen}[1]{{\uparrow_{#1}}}
\newcommand{\total}{\theta}
\newcommand{\Total}{\Theta}
\newcommand{\Partial}{\Delta}
\newcommand{\switch}{\ensuremath{\mathsf{switch}}\xspace}
\newcommand{\resp}{\ensuremath{\mathsf{resp}}\xspace}
\newcommand{\blame}{\ensuremath{\mathsf{blame}}\xspace}
\newcommand{\reduce}{\ensuremath{\mathsf{DLS}}\xspace}
\newcommand{\thresh}{\ensuremath{\tau}\xspace}

\newcommand{\profeat}{\textsc{ProFeat}}
\newcommand{\qflan}{\textsc{QFLan}}
\newcommand{\featcause}{\textsc{FeatCause}}
\newcommand{\espresso}{\textsc{Espresso}}
\newcommand{\signature}{\textsc{Espresso-Signature}}
\newcommand{\provelines}{\textsc{ProVeLines}}
\newcommand{\splverifier}{\textsc{SPLVerifier}}
\newcommand{\pyeda}{\textsc{PyEDA}}
\newcommand{\simulink}{\textsc{SimuLink}}

\newcommand{\minepump}{\textsc{Minepump}\xspace}
\newcommand{\elevator}{\textsc{Elevator}\xspace}
\newcommand{\cfdp}{\textsc{CFDP}\xspace}
\newcommand{\Email}{\textsc{E\-mail}\xspace}

\newcommand{\llvm}{\textsc{LLVM}\xspace}
\newcommand{\lrzip}{\textsc{Lrzip}\xspace}
\newcommand{\dune}{\textsc{DUNE}\xspace}
\newcommand{\berkeley}{\textsc{BerkeleyDB}\xspace}
\newcommand{\xtwosixfour}{\textsc{x264}\xspace}
\newcommand{\apache}{\textsc{Apache}\xspace}
\newcommand{\linux}{\textsc{Linux}\xspace}
\newcommand{\SQLite}{\textsc{SQLite}\xspace}
\newcommand{\wget}{\textsc{WGet}\xspace}
\newcommand{\tvl}{\textsc{TVL}\xspace}

\copyrightyear{2022} 
\acmYear{2022} 
\setcopyright{acmcopyright}\acmConference[ICSE '22]{44th International Conference on Software Engineering}{May 21--29, 2022}{Pittsburgh, PA, USA}
\acmBooktitle{44th International Conference on Software Engineering (ICSE '22), May 21--29, 2022, Pittsburgh, PA, USA}
\acmPrice{15.00}
\acmDOI{10.1145/3510003.3510200}
\acmISBN{978-1-4503-9221-1/22/05}

\title{Causality in Configurable Software Systems}

\author{Clemens Dubslaff}
\email{clemens.dubslaff@tu-dresden.de}
\affiliation{%
    \institution{Centre for Tactile Internet with Human-in-the-Loop (CeTI)\\
    Technische Universit\"at Dresden}
    \city{Dresden}
    \country{Germany}
  }
  
\author{Kallistos Weis}
\email{kallistos@cs.uni-saarland.de}
\affiliation{%
    \institution{Saarland University\\
    Saarland Informatics Campus}
    \city{Saarbrücken}
    \country{Germany}
  }
  
\author{Christel Baier}
\email{christel.baier@tu-dresden.de}
\affiliation{%
    \institution{Technische Universit\"at Dresden}
    \city{Dresden}
    \country{Germany}
  }
  
\author{Sven Apel}
\email{apel@cs.uni-saarland.de}
\affiliation{%
    \institution{Saarland University\\
    Saarland Informatics Campus}
    \city{Saarbrücken}
    \country{Germany}
  }

\begin{abstract}
	Detecting and understanding reasons for defects and inadvertent behavior
	in software is challenging due to their increasing complexity. 
	In configurable software systems, the combinatorics that
	arises from the multitude of features a user might select from
	adds a further layer of complexity.
	We introduce the notion of \emph{feature causality}, which
	is based on \emph{counterfactual reasoning} and inspired by the
	seminal definition of actual causality by Halpern and Pearl.
	Feature causality operates at the level of system configurations and
	is capable of identifying features and their interactions that are the 
	reason for emerging functional and non-functional properties.
	We present various methods to explicate these reasons, in particular
	well-established notions of \emph{responsibility} and \emph{blame}
	that we extend to the feature-oriented setting.
	Establishing a close connection of feature causality to prime implicants,
	we provide algorithms to effectively compute feature causes and causal explications.
	By means of an evaluation on a wide
	range of configurable software systems, including community benchmarks and
	real-world systems, we demonstrate the feasibility of our approach:
	We illustrate how our notion of causality facilitates to identify root causes, 
	estimate the effects of features, and detect feature interactions.
\end{abstract}

\setcopyright{none} 

\begin{document}
\maketitle

\section{Introduction}\label{sec:introduction}
Configurable software systems offer a wide variety of configuration options that control
the activation of features desired by the user
and that influence critical functional and non-functional properties such as performance.
The often huge configuration spaces render the detection, prediction, and explanation
of defects and inadvertent behavior challenging tasks.
While there are specifically tailored analysis methods to tackle this
challenge~\cite{Thum14}, research on their \emph{explainability} is still in its infancy~\cite{BaiDubFun21}.
The potentially exponential number of system configurations and corresponding
analysis results, bug reports, or other feature-dependent properties
demand techniques for a meaningful and feasible interpretation.

In this paper, we present a set of fundamental concepts and methods to identify and interpret
properties of configurable systems \emph{at the level of features} by \emph{causal reasoning}.
We introduce the notion of \emph{feature causes} as feature selections that
are the reason for emergent system behaviors. Our notion of feature causality is inspired 
by the seminal counterfactual definition of \emph{actual causality} by Halpern and 
Pearl~\cite{HalPearl01-Causes,Halpern2015}.
Relevant analysis and reasoning tasks that we address include to determine
features that cause a bug, the degree to which some configurations are responsible for
bad system performance, or which features necessarily have to interact for
inadvertent behavior.

Since features correspond to system functionalities specified
by software engineers, they often have a dedicated meaning in the target 
application domain~\cite{ApeBatKas13a}.
To this end, defects (and other behaviors of interest) detected at the level
of features can provide important insights for the resolution of
\emph{variability bugs}~\cite{GarCoh11a,Rhein2018,Abal2018} and
\emph{configuration-dependent behavior}~\cite{SieKolKaeApe2012,SieGreApe15a,GYS+18,NYM+20}.
As such, they are certainly more informative and actionable than low-level
program traces alone.
Developers may choose to focus on those feature implementations identified as
root causes of bugs or simply disallow or coordinate the activation of certain
features when defects are related to them.

Our presented techniques are generic in the sense that they are neither language-,
architecture-, nor environment-specific and applicable within any effective method to
analyze or test variability-aware properties.
To this end, causal reasoning on both variability-aware white-box and 
black-box analyses is supported.
This complements existing causal reasoning techniques for the detection of root causes:
Approaches such as delta-debugging~\cite{Zel02a,CleZel05a},
causal testing~\cite{JohBruMel20}, or causal trace analysis~\cite{BeerBCOT2012}
require a white-box analysis that operates at the level of code and are not variability-aware.
Hence, they usually would have to be applied on a multitude of system
configurations for a variability-aware causal analysis, suffering from a
combinatorial blowup.
We envision applications of feature causality at those development phases where
analysis methods are used, for instance, in software product line engineering.
Also in production-level deployments our techniques shall be useful to optimize
software through causally relevant configurations.

\tudparagraph{Evaluation}
We present algorithms to compute feature causes, represent, and interpret them
by means of concise logic formulas, feature interactions, and
\emph{responsibility} and \emph{blame}~\cite{ChoHal04a}.
Our prototypical implementation relies on \emph{binary decision diagrams (BDDs)}~\cite{Bryant86}
and the computation of prime implicants using the de-facto
standard two-level logic minimizer \espresso{}~\cite{McGSanBra93a}.
By means of an analysis of several configurable systems, including community benchmarks and
real-world systems, we investigate feature causes and their properties.
We demonstrate that our notion of feature causes and methods to represent them help to
pinpoint features relevant for the configurable system's properties and
illustrate how feature interactions can be detected and quantified.

\tudparagraph{Contributions}
In summary, our contributions are:
\\[.3em]
\begin{tabular}{p{.001\columnwidth}p{.94\columnwidth}}
	$\bullet$ & We introduce the notion of feature causality inspired by the well-established counterfactual
		definition of actual causality by Halpern and Pearl~\cite{HalPearl01-Causes,Halpern2015}.\\
	$\bullet$ & We show that feature causes coincide with certain prime implicants,
		leading to an algorithm to compute all feature causes.\\
	$\bullet$ & We provide methods to interpret and represent feature causes by propositional formulas,
		responsibility and blame, and potential feature interactions.\\
	$\bullet$ & We offer a BDD-based prototype to compute and represent
		feature causes and feature interactions.\\
	$\bullet$ & We conduct several experiments illustrating how to determine and reason
		about feature causes in different realistic settings.
\end{tabular}~\\[-.5em]

\tudparagraph{Supplement}
Proofs of the theoretical statements are included in \cite{Dub21}.
The source code of our implementation and raw data to reproduce our 
experiments are publicly available \cite{DubWeiBai22,featcause}.

\section{Background}\label{sec:background}
In this section, we revisit basic concepts and notions
from logics and configurable systems used throughout the paper.

\tudparagraph{Interpretations}
A \emph{partial interpretation} over a set $X$ is a partial mapping
$\partial\colon X \pa \{\true,\false\}$. We denote by $\supp(\partial)$ the
\emph{support} of $\partial$, i.e., the set of all elements $x\in X$ where $\partial(x)$ is
defined. We say that $\partial$ is a \emph{total interpretation} if $\supp(\partial)=X$
and denote by $\Partial(X)$ and $\Total(X)$ the set of partial and total
interpretations, respectively.
Given a partial interpretation $\partial\in\Partial(X)$, we define its
semantics $\Sem{\partial}\subseteq\Total(X)$ as the set of all total
interpretations $\total\in\Total(X)$ where for all $x\in\supp(\partial)$
we have $\partial(x)=\total(x)$. 
We say that $\partial\in\Partial(X)$ \emph{covers} $\partial'\in\Partial(X)$
if $\Sem{\partial'}\subseteq\Sem{\partial}$.
For a set of partial interpretations
$\cP\subseteq\Partial(X)$, we define $\Sem{\cP}=\bigcup_{\partial\in\cP} \Sem{\partial}$.
The \emph{$x$-expansion} of a partial interpretation $\partial\in\Partial(X)$
is the partial interpretation $\partial\gen{x}\in\Partial(X)$ where 
$\supp(\partial\gen{x})=\supp(\partial){\setminus} \{x\}$
and where $\partial\gen{x}(y)=\partial(y)$ for all $y\in \supp(\partial){\setminus} \{x\}$.
For a set of total interpretations $\cT\subseteq\Total(X)$, an \emph{implicant} of
$\cT$ is a partial interpretation $\partial\in\Partial(X)$ where
$\Sem{\partial}\subseteq\cT$. We call $\partial$ a \emph{prime implicant}
if $\partial$ is minimal, i.e., $\Sem{\partial\gen{x}}\not\subseteq\cT$
for all $x\in\supp(\partial)$.

\tudparagraph{Propositional logics}
A \emph{propositional logic formula} over a set $X$ is an expression defined by the grammar
\begin{center}
	$\phi \en{=} \true \mid \false \mid x \mid \neg\phi \mid
		\phi\wedge\phi \mid \phi\vee\phi$
\end{center}
where $x$ ranges over $X$.
The \emph{length} $|\phi|$ of a formula $\phi$ is recursively
defined by $|\true| =|\false| =|x| =1$, $|\neg\phi| =|\phi|+1$, and
$|\phi_0\wedge\phi_1| =|\phi_0\vee\phi_1| =|\phi_0|+|\phi_1|+1$.
For a partial interpretation $\partial\in\Partial(X)$, we write $\partial\models\phi$
if either $\phi = \true$, $\phi = x \in\supp(\partial)$ and $\partial(x)=\true$,
$\phi = \neg\psi$ and not $\partial\models\psi$, $\phi = \phi_0 \wedge \phi_1$ and
$\partial\models\phi_0$ and $\partial\models\phi_1$, and $\phi = \phi_0 \vee \phi_1$ and
$\partial\models\phi_0$ or $\partial\models\phi_1$. The semantics of $\phi$ is
the set of all satisfying total interpretations $\Sem{\phi}= \{ \total\in\Total(X) : \total\models\phi\}$.

\tudparagraph{Configurable systems}
A widely adopted concept to model configurable systems is by means of
features~\cite{ApeBatKas13a}.
\emph{Features} encapsulate optional or incremental units of functionality~\cite{Zave2001}
and are used to describe commonalities and variabilities of whole families of systems.
At an abstract level, we fix a set $\Feat$ of Boolean configuration options where
each element corresponds to some feature of the system.
We call a total interpretation $\total\in\Total(\Feat)$ over $\Feat$ a \emph{configuration},
which we usually describe by listing the selected features, i.e.,
the features $x\in\Feat$ where $\total(x)=\true$.
The set of \emph{Valid configurations} $\ValidFeat\subseteq\Total(\Feat)$ comprises
those configurations for which there exists a corresponding system implementation,
usually specified by a \emph{feature diagram}~\cite{Kang1990}.
A partial interpretation over $\Feat$ is called \emph{partial configuration}.
	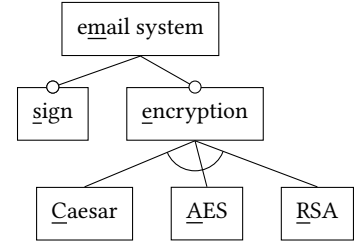
\begin{figure}[t]
		\centering%

\begin{tikzpicture}[featurediagram]

  \tikzset{features/.style={%
      matrix of nodes,nodes=feature
    }
  }

  \node[feature] (root) {e\underline{m}ail system};

  \matrix(sub)[features,below=3mm of root,column sep=.5cm] {%
  \underline{s}ign & \underline{e}ncryption\\
  };

  \matrix(enc)[features,below=of sub-1-2,column sep=.5cm] {%
    \underline{C}aesar & \underline{A}ES & \underline{R}SA\\
  };

  \xordecomp{sub-1-2}{enc-1-1}{enc-1-3}
  
  \draw[optional] (sub-1-1.north) circle (.8mm);

  \path (root.south)
    edge (sub-1-1.north)
    edge (sub-1-2.north);
    
  \draw[optional] (sub-1-1.north) circle (.8mm);
  \draw[optional] (sub-1-2.north) circle (.8mm);

  \path (sub-1-2.south) %
    edge (enc-1-1.north)
    edge (enc-1-2.north)
    edge (enc-1-3.north);
    
\end{tikzpicture}
		\Description{Feature diagram for the email system example}
		\caption{\label{fig:fd-mail}Feature diagram for the email system example}
	\end{figure}
\begin{example}\label{bsp:email}
	As the running example,	consider a simple email system over features $F=\{m,s,e,c,a,r\}$,
	formalizing the base e\underline{m}ail system functionality,
	optional features for \underline{s}igning and \underline{e}ncryption, 
	and encryption methods \underline{C}aesar, \underline{A}ES,
	and \underline{R}SA.
        For the encryption features, we assume that exactly
	one can be selected. The described variability constraints for the email system
	are specified in the feature diagram shown in Figure~\ref{fig:fd-mail},
	leading to valid configurations
\[
	\ValidFeat \en{=}
		\{\, m, mec, mea, mer, ms, msec, msea, mser\,\}.
\]
\end{example}

\section{Feature Causality}\label{sec:causality}
The notion of \emph{causality} has been extensively studied in philosophy,
social sciences, and artificial intelligence~\cite{Good59,Eells91,Pearl09,Wil09}.
We focus on \emph{actual causality}, describing
binary causal relationships between cause events $C$ and effect events $E$.
Halpern and Pearl formalized actual causality based on the concept of
\emph{counterfactual dependencies}~\cite{Lewis73} using a structural-equation
approach~\cite{Halpern2015,HalPearl01-Causes,HalPearl01-Explanations}.
The idea of counterfactual reasoning \cite{WacMitRus17a} relies on the assumption
that $E$ would not have happened if $C$ had not happened before, which corresponds
to the standard ``but-for'' test used in law.

In this section, we take inspiration of the definition by Halpern and
Pearl~\cite{HalPearl01-Causes,Halpern2015} to establish a notion of causality
at the level of features.
Here, we interpret the selection of features as events considered for actual causality.
The basic reasoning task we address then amounts to 
\emph{determine those feature selections that cause a given effect property}.
Examples for effect properties are ``the execution time is longer than five minutes''
or ``the system crashes''.

We assume to have described the effect properties as 
\emph{effect set} $\Effects\subseteq\ValidFeat$ of valid
configurations for which an effect property can be observed.
Elements of $\Effects$ are called \emph{effect instances}.
All other valid configurations in $\ValidFeat{\setminus}\Effects$ are
assumed not to exhibit the effect.
Feature selections are naturally specified by partial configurations.
Clearly, a partial configuration $\gamma$ can only be a cause of the
effect if $\gamma$ ensures the effect to emerge, i.e.,
all valid configurations covered by $\gamma$ are effect instances. 
Furthermore, following counterfactual
reasoning, we require for $\gamma$ being a cause that, if we would select features
of $\gamma$ differently, there might be a configuration for which the effect does
not emerge. These two intuitive conditions on causality
are reflected in our formal definition of causes of $\Effects$ w.r.t. $\ValidFeat$:

\begin{definition}\label{def:fc}
A \emph{feature cause} of an effect $\Effects$ w.r.t. valid
configurations $\ValidFeat$ is a partial configuration $\gamma\in\Partial(\Feat)$ where
\begin{enumerate}[label=\textbf{(FC$_\arabic*$)},leftmargin=*,ref=\textbf{FC$_\arabic*$}]
	\item\label{en:fc1} 	$\varnothing\neq\Sem{\gamma}\cap\ValidFeat\subseteq\Effects$ and
	\item\label{en:fc2}
			$\Sem{\gamma\gen{x}}\cap\ValidFeat \not\subseteq \Effects$ for all $x\in\supp(\gamma)$.
\end{enumerate}
We denote by $\Causes$ the set of all causes for $\Effects$ w.r.t.\ $\ValidFeat$.
\end{definition}
In case \ref{en:fc1} holds for a partial configuration $\gamma$, we say that
$\gamma$ is \emph{sufficient} for $\Effects$ w.r.t.\ $\ValidFeat$~\cite{GarCoh11a}.
The counterfactual nature of \ref{en:fc2} ensures that for every feature cause $\gamma$ 
and $x\in\supp(\gamma)$ there is a \emph{counterfactual witness}
$\overline{\eta}\in\Sem{\gamma\gen{x}}\cap(\ValidFeat{\setminus}\Effects)$.
That is, a valid feature configuration where the effect does not emerge but where
changing one feature selection may yield an effect instance.
Note that \ref{en:fc2} ensures minimality of the feature cause w.r.t. its support,
i.e., dropping conditions on interpretations of features necessarily leads to
a partial configuration that is not sufficient for the effect anymore.
In the formal definition of Halpern and Pearl causality~\cite{Halpern2015}, 
counterfactuality and minimality are stated in two distinct conditions.

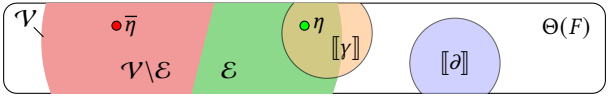
\begin{figure}[t]%
	\centering%
\definecolor{dgreen}{rgb}{0.1, 0.7, 0.1}
\definecolor{dred}{rgb}{0.9, 0.2, 0.2}

\def\cecircle{(2.5cm,.7cm) circle (2cm)}
\def\ecircle{(4cm,.7cm) circle (5cm)}
\def\gcircle{(4.3cm,.8cm) circle (.6cm)}
\def\pcircle{(6cm,.4cm) circle (.6cm)}
\def\trect{(0,0) rectangle ++(8cm,1.2cm)}

\begin{tikzpicture}
	\draw[rounded corners=5pt] \trect node[below left=.1cm] {$\Total(\Feat)$};

	\begin{scope}
		\clip \trect;
		\fill[dred!50] \cecircle;
	\end{scope}

	\begin{scope}
		\clip \trect;
		\clip \cecircle;
		\clip (2cm,-2cm) -- (3cm,2cm) -- (8cm,2cm) -- (8cm,-2cm) -- cycle;
		\fill[dgreen!50] \ecircle;
	\end{scope}
	
	\draw (3cm,.3cm) node {$\Effects$};
	\draw (1.9cm,.29cm) node {$\CEffects$};
	\draw (.3cm,1cm) node {$\ValidFeat$};
	\draw (.4cm,.88cm) -- (.53cm,.75cm);

	\begin{scope}
		\clip \trect;
		\draw[fill=orange!50, opacity=.6] \gcircle node[below right=-.1cm, opacity=1] {$\Sem{\gamma}$};
	\end{scope}
	\draw[fill=red] (1.5cm,.9cm) circle (.06cm) node[right] {$\overline{\eta}$};
	\draw[fill=green] (4cm,.9cm) circle (.06cm) node[right] {$\eta$};

	\begin{scope}
		\clip \trect;
		\draw[fill=blue!30, opacity=.6] \pcircle node[opacity=1] {$\Sem{\partial}$};
	\end{scope}

\end{tikzpicture} %
	\Description{Feature configuration sets for feature causality}%
	\caption{\label{fig:sets}Configuration sets for feature causality}%
\end{figure}%
We usually denote configurations in $\Effects$ by $\eta$, counterfactual
witnesses in $\CEffects$ by $\overline{\eta}$, and feature causes by $\gamma$.
Figure~\ref{fig:sets} depicts the relation between valid configurations, 
effects, causes, and counterfactual witnesses.

\begin{example}\label{bsp:emailcause}
	Let us continue our running example of the configurable email system
	introduced in Example~\ref{bsp:email}.
    We consider an effect property reflecting ``long decipher time'', e.g., that it takes
	in average more than three months for an attacker to decrypt an email.
	Assume that this effect property can be observed by configurations
	in which AES or RSA are selected, i.e., $\Effects = \{ mea, mer, msea, mser\}$.
    Conversely, in all valid configurations in which AES and RSA are not selected,
	the effect does not emerge.
	In this setting, the encryption features AES and RSA are both causes since
	all valid configurations with either feature show the effect.
	Considered in isolation, AES and RSA are not necessary for the effect, as 
	one can choose the other encryption feature (RSA or AES, respectively) to ensure the effect.
	The sign feature does not trigger the effect and is not a cause.
	
	Interestingly, a further cause is given by selecting the encryption feature and explicitly
	deselecting the Caesar feature, illustrating that also explicitly not selecting
	features might be a cause of some effect.
	This hints at the fact that causes can be represented
	in different ways, addressed later in the paper.
	
	Formalizing this intuition, we check whether the three partial configurations
	$\gamma_a$, $\gamma_r$, and $\gamma_{e\bar{c}}$ given by
	\begin{enumerate}[label=(\roman*),leftmargin=*]
		\item $\supp(\gamma_a)=\{a\}$ with $\gamma_a(a)=\true$,
		\item $\supp(\gamma_r)=\{r\}$ with $\gamma_r(r)=\true$, and
		\item\label{it:desfc} $\supp(\gamma_{e\bar{c}})=\{e,c\}$ with $\gamma_{e\bar{c}}(e)=\true$
			and $\gamma_{e\bar{c}}(c)=\false$
	\end{enumerate}
	are indeed feature causes of $\Effects$ w.r.t. $\ValidFeat$ according to 
	Definition \ref{def:fc}: First, note that 
	$\Sem{\gamma_{e\bar{c}}}\cap\ValidFeat = (\Sem{\gamma_a}\cup\Sem{\gamma_r})\cap\ValidFeat = \Effects$
	and both, $\Sem{\gamma_a}\cap\ValidFeat$ and $\Sem{\gamma_r}\cap\ValidFeat$, are non-empty.
	Hence, \ref{en:fc1} is fulfilled for $\gamma_{e\bar{c}}$, $\gamma_a$, and $\gamma_r$.
	To check \ref{en:fc2}, we observe that 
	$\Sem{\gamma_a\gen{a}} = \Sem{\gamma_r\gen{r}} = \Total(\Feat)$	and
	\begin{eqnarray*}
		\Sem{\gamma_{e\bar{c}}\gen{e}} \cap\ValidFeat &=& \Effects\cup\{m, ms\},\text{ and}\\
		\Sem{\gamma_{e\bar{c}}\gen{c}} \cap\ValidFeat &=& \Effects\cup\{mec, msec\}.
	\end{eqnarray*}
	Hence, $m$ is a counterfactual witness for $\gamma_a$, $\gamma_r$, and $\gamma_{e\bar{c}}$
	w.r.t. $a$, $r$, and $e$, respectively, while $mec$ can serve as such for $\gamma_{e\bar{c}}$
	and $c$. It is easy to check that there are no further feature causes since for all other 
	partial 	configurations sufficient for $\Effects$ w.r.t. $\ValidFeat$ (\ref{en:fc1})
	there are expansions towards $\gamma_a$, $\gamma_r$, or $\gamma_{e\bar{c}}$ 
	(thus, violating \ref{en:fc2}).
	
\end{example}

\subsection{Effect Properties and Sets}\label{sec:effects}
Our definition of feature causality relies on a given effect set, which is assumed
to comprise all those valid configurations where the effect property holds.
We now elaborate more on how to obtain effect sets
from analyzing configurable systems. 
In fact, our generic definition supports
a multitude of effect properties for which the only assumption is that there is an 
effective method to determine all configurations in which the effect property holds. 
Such methods also include variability-aware white-box analyses~\cite{WAS21,VJS+21},
where the source code or operational behavior of system variants is accessible,
as well as black-box analyses relying on testing or sampling~\cite{GYS+18,KalGreSie2019}.
In the following paragraphs, we exemplify how to obtain effect sets from
analysis results. Our discussions reflect the effect properties in the experimental 
evaluation section (see Section~\ref{sec:casestudy}) and do not claim to be exhaustive.

\tudparagraph{Functional properties}\label{sec:functional}
To reason about causality w.r.t.\ functional properties, the effect set can
be determined by variability-aware static analysis~\cite{BeeDamLie19a,Rhein2018}
or model checking~\cite{PlaRya2001,ClaCorSchobHeyLegRas13,ApeRheWen2013}.
In the latter case, %
effect properties can be formalized, e.g., in a temporal logic such as
LTL~\cite{Pnu77a} or CTL~\cite{ClaEmeSis1986}.
Model checking configurable systems against LTL and CTL properties has broad tool 
support~\cite[e.g.][]{Classen2012,ProveLines13}. 
Given a formula $\varphi$ that specifies the effect property, 
these tools explicitly return the effect set
\[
	\Effects_\varphi \en{=}
	\{ \total \in\ValidFeat : \total\models\varphi\}
\]
of valid configurations $\total\in\ValidFeat$ whose corresponding
system variants satisfy $\varphi$.
Since model checking is based on an exhaustive analysis, an analysis also exposes those
valid configurations for which the effect property does not hold.
The same is possible for variability-aware static analysis~\cite{Rhein2018,BodTolRib13}.

\tudparagraph{Non-functional properties}\label{sec:nonfunctional}
Besides functional properties, also non-functional properties of configurable
systems can serve as effect property and give rise to an effect set.
Let $\rho\colon \ValidFeat \ra \Real$ be a function that results from a quantitative
analysis of the configurable system in question. Values $\rho(\total)$ for a valid configuration
$\total\in\ValidFeat$ may stand for the performance achieved, the probability
of failure, or the energy consumed in the system variant that corresponds to $\total$.
To obtain $\rho$ for real-world systems, Siegmund et al.~\cite{SieGreApe15a, SieKolKaeApe2012}
presented a black-box method to generate linear-equation models for
performance measures by multivariable linear regression on sampled configurations.
Other black-box approaches rely on regression trees~\cite{GYS+18}, Fourier
learning~\cite{ZhaGuoBla+15}, or probabilistic programming~\cite{DorApeSie20}.
Related white-box approaches use insights of local measurements and taint analysis 
information~\cite{VJS+21} or profiling information~\cite{WAS21}.
Variability-aware probabilistic model checking pursues a white-box analysis approach
on state-based operational models where effect properties are
specified in quantitative variants of temporal logic~\cite{DBK15,BeeLegLlu16a}.
These approaches have been implemented in the tools $\profeat$~\cite{CDKB18} 
and $\qflan$~\cite{VanBeeLeg18a}.

Given $\rho$ that results from one of the analysis
approaches mentioned above, an effect set can be specified by imposing a
threshold $\thresh\in\Real$ combined with a comparison relation $\sim$ towards
\[
	\Effects_{\rho\sim\thresh} \en{=}
	\{ \total \in\ValidFeat : \rho(\total)\sim\thresh\}.
\]
\begin{example}\label{bsp:effectset}
	In Example~\ref{bsp:emailcause}, we informally specified the effect
	of a ``long decipher time'' as taking more than three months to decrypt
	an email without having the encryption key available. 
	By a variability-aware quantitative analysis on the email system,
    we may obtain a function $\rho$ that,
    for a configuration $\total$, returns the minimal time in years to decipher
    an email sent with the system variant corresponding to $\total$.
	Analysis results could be, e.g., $\rho(\total)=0$ with no encryption,
	$\rho(\total)=10^{-7}$ with Caesar, $\rho(\total)=1$ with AES, and $\rho(\total)=2$
	with RSA selected in $\total$, respectively.
	Then, $\Effects_{\rho>0.25}$ provides the effect set
	$\Effects$ of Example~\ref{bsp:emailcause}.
\end{example}

\tudparagraph{On computing effect sets}
The effect set and the set of valid configurations can be of exponential size
in the number of features. An efficient computation of these sets depends on the
analysis techniques used and are independent from our causal framework.
However, specifically tailored variability-aware analysis techniques can tackle
the exponential blowup, e.g., through symbolic representation of family models~\cite{Thum14,Dub19a}.

\subsection{Computation of Feature Causes}\label{sec:algorithm}
For a given effect set $\Effects$ and a set of valid configurations $\ValidFeat$
along with a partial configuration $\partial$, Definition~\ref{def:fc} directly
provides a polynomial-time algorithm to decide whether $\partial$ 
is a cause of $\Effects$ w.r.t. $\ValidFeat$ by checking \ref{en:fc1} and \ref{en:fc2}.
From this, we obtain a simple approach to compute the set $\Causes$ by successively
checking expansions for sets of features applied on elements in $\Effects$ as candidates 
for causes. Since there might be exponentially many such expansions, this approach 
easily renders infeasible already within a small number of features.

In this section, we present a practical algorithm to compute the set of causes,
which relies on a connection of Definition~\ref{def:fc} to the notion
of prime implicants (see Section~\ref{sec:background}):

\begin{lemma}\label{lem:primecomp}
	For any partial configuration $\partial\in\Partial(\Feat)$
\[
		\Sem{\partial}\cap\ValidFeat\subseteq\Effects \en{\text{ iff }}
		\Sem{\partial}\subseteq \big(\Total(\Feat){\setminus}\ValidFeat\big)\cup\Effects.
\]
\end{lemma}

Following this lemma, every cause of $\Effects$ w.r.t. $\ValidFeat$
is also an implicant of $\big(\Total(\Feat){\setminus}\ValidFeat\big) \cup \Effects$
due to \ref{en:fc1} and even a prime implicant due to \ref{en:fc2}.
Conversely, every prime implicant $\partial$ of 
$\big(\Total(\Feat){\setminus}\ValidFeat\big) \cup \Effects$ for which 
$\Sem{\partial}\cap\Effects\neq\varnothing$ is a cause due to \ref{en:fc1}.
This directly suggests an algorithm to compute causes via prime implicants:
\begin{algorithm}[t]\small
	\SetAlgoLined
	\DontPrintSemicolon
	\SetKwInOut{Input}{input}\SetKwInOut{Output}{output}
	\Input{$\Effects,\ValidFeat\subseteq\Total(\Feat)$}\Output{$\Causes$}
	\BlankLine
	\lIf{$\Effects=\varnothing$}{\Return $\varnothing$\label{l:test}}
	$\cP:=\textsc{Compute-Primes}\big((\Total(\Feat){\setminus}\ValidFeat)\cup\Effects\big)$\label{l:prime}\;
	\lForAll{$\partial\in \cP$ where $\Sem{\partial}\cap\Effects = \varnothing$}{$\cP := \cP{\setminus}\{\partial\}$%
			\label{l:remove}}%
	\Return $\cP$ \;
	\Description{Computation of feature causes via prime implicants}
	\caption{\label{algo:primecause}Computation of feature causes}
\end{algorithm}
Algorithm~\ref{algo:primecause} first generates prime implicants as cause
candidates and then removes those candidates that are not sufficient for
$\Effects$ w.r.t.\ $\ValidFeat$.
Figure~\ref{fig:sets} reflects this situation where
$\gamma$ and $\partial$ are prime implicants with
$\gamma$ being a cause and $\partial$ not: at least one effect instance
is covered by $\gamma$, while this is not the case 
for $\partial$ and hence would be removed by Algorithm~\ref{algo:primecause}.
Prime implicants of a set of configurations can be computed in polynomial time in
the size of the input set~\cite{Str92a}, which directly leads to:
\begin{theorem}\label{thm:computecauses}
	Given valid configurations $\ValidFeat\subseteq\Total(\Feat)$ 
	and effect set $\Effects\subseteq\ValidFeat$,
	Algorithm~\ref{algo:primecause} computes $\Causes$, the set of feature causes for
	$\Effects$ w.r.t. $\ValidFeat$, in polynomial time in $|\Total(\Feat)|$.
\end{theorem}
Note that the set of valid configurations and the effect set can be both exponential in the
number of features and there might be exponentially many prime implicants~\cite{ChaMar78a} 
in the worst case. Hence, Algorithm~\ref{algo:primecause} is exponential 
in the number of features.

\begin{example}
	Let us illustrate the computation of feature causes of Example~\ref{bsp:emailcause}
	by Algorithm~\ref{algo:primecause}. First notice that
	\[
	\big(\Total(\Feat)\setminus\ValidFeat\big)\cup\Effects \en{=}
		\Total(\Feat)\setminus\{\, m, mec, ms, msec \,\}
	\]
	comprising $60$ feature configurations. 
	The prime implicants for this set are computed in Line~\ref{l:prime}, which yields
	\[
		\cP \en{=} \{\,\gamma_{\bar{m}}, \gamma_a, \gamma_r, \gamma_{e\bar{c}}, \gamma_{\bar{e}c}\,\}.
	\]
	Here, we used notations as in Example~\ref{bsp:emailcause}, e.g., $\supp(\gamma_{\bar{e}c})=\{e,c\}$,
	$\gamma_{\bar{e}c}(e)=\false$, and $\gamma_{\bar{e}c}(c)=\true$.
	Clearly, all configurations covered by $\gamma_{\bar{m}}$ or $\gamma_{\bar{e}c}$ are
	not valid and hence also no effects. Thus, they are removed in Line~\ref{l:remove}, leading
	to $\Causes=\{\gamma_a, \gamma_r, \gamma_{e\bar{c}}\}$.
	
\end{example}

\section{Causal Explications}\label{sec:minimal}
Since the number of feature causes can be exponential in the number of features,
a mere listing of all causes is neither feasible nor expedient
for real-world software systems.
This holds for both, humans that have to evaluate causal
relationships in configurable systems, e.g., during software development,
and machines that might use feature causes for further processing and reasoning.

In this section, we present and discuss several methods to compute \emph{causal explications},
i.e., mathematical or computational constructs that arise from processing
feature causes to provide useful causal representations and measures~\cite{BaiDubFun21}.
Explications are closely related to \emph{explanations},
by which we mean human-understandable objects employed within an integrated system, 
e.g., in feature-oriented software development or in production-level deployments.

Our methods for computing explications rely on techniques from propositional logic and
circuit optimization~\cite{Pau75a,McGSanBra93a}, responsibility and blame~\cite{ChoHal04a},
and feature interactions~\cite{GarCoh11a}.
They all take a global perspective on sets of feature causes
rather than only considering single feature causes in isolation.
In the following, we fix sets of valid configurations
$\ValidFeat\subseteq\Total(\Feat)$ and effects $\Effects\subseteq\ValidFeat$.
\subsection{Distributive Law Simplification}\label{sec:propositional}
A rather natural explication for a set of causes $\cC\subseteq\Causes$ is to
represent $\cC$ as propositional logic formula, e.g.,
as the \emph{characteristic formula}
$\chi(\cC)$ defined in disjunctive normal form (DNF)
\[
	\chi(\cC) \en{=}
		\bigvee_{\partial\in\cC} \big(
			\bigwedge_{x\in\supp(\partial) \atop \partial(x)=\true} x \wedge
			\bigwedge_{x\in\supp(\partial) \atop \partial(x)=\false} \neg x
		\quad\big).
\]
Clearly, $\chi(\cC)$ has the same size as $\cC$ and its representation does not
exhibit any advantage compared to $\cC$. 
Methods to minimize propositional logic formulas~\cite{McC56a,McGSanBra93a,HemSch11a}
could be used to yield small formulas $\varphi$ covering the same configurations as $\cC$, 
i.e., where $\Sem{\varphi}=\Sem{\cC}$.
While beneficial for related problems in configurable systems analysis, e.g., for 
presence condition simplification~\cite{RheGreApe15a}, such methods 
vanish causal information, i.e., the set of causes $\cC$ cannot be reconstructed from 
the reduced formula $\varphi$.
To provide a small formula that maintains the causal information
of $\cC$, we use a simple yet effective reduction method, which we call
\emph{distributive law simplification (DLS)}.
The basic idea is to factorize common feature selections in a DNF
formula step by step, exploiting the $n$-ary distributive law
$\Sem{\bigvee_{i=1}^n (\phi\wedge \psi_i)} = \Sem{\phi \wedge \bigvee_{i=1}^n\psi_i}$.
Each factorization leads to a length reduction of $(n-1)\cdot|\phi|$, where
$|\phi|$ is the length of the propositional formula factored out.
Obviously, these transformations are reversible, such that the original
DNF $\chi(\cC)$ and hence the set of causes $\cC$ can be reconstructed.
The final formula length depends on the formulas factored out, the subformulas,
and the factorization order. Determining a formula through DLS that has minimal size 
is close to global optimization problems for propositional logic formulas and thus
computationally hard.
For practical applications, we hence employ a heuristics that reduces a given formula $\varphi$ 
in DNF by stepwise factoring out literals that have maximal number of occurrences 
in DNF subformulas. We denote the reduced formula obtained by this heuristics by $\reduce(\varphi)$.
\subsection{Cause--Effect Covers}\label{sec:smin}
The complete set of causes may contain several candidates to describe reasons for the effect
emerging in a single system variant.
If not interested in all causes but in a set of causes that covers
all effects (i.e., that contains, at least, one cause for every system variant)
we might ask for a preferably small \emph{covering set}.
Formally, a \emph{cause--effect cover} of $\Effects$ is a set $\cC\subseteq\Causes$
where $\Effects\subseteq\Sem{\cC}$. We say that such a cover $\cC$ is
\emph{minimal} if there is no cause--effect cover $\cC'$ of $\Effects$
where $|\cC'|<|\cC|$.

\begin{example}
For the email system in Example~\ref{bsp:emailcause} we directly see that $\{\gamma_{e\bar{c}}\}$
and $\{\gamma_a,\gamma_r\}$ are the only cause--effect covers of $\Effects$.
Thus, $\{\gamma_{e\bar{c}}\}$ is a minimal cause--effect cover of $\Effects$
as $|\{\gamma_a,\gamma_r\}|>|\{\gamma_{e\bar{c}}\}|$.
\end{example}

Determining minimal cause--effect covers is computationally hard for
the same reasons as for minimal prime--cover computation~\cite{Pau75a}.
Hence, for practical applicability, heuristics that lead to nearly
minimal cause--effect covers are of interest.

\begin{definition}\label{def:maxfc}
The binary relation $\unlhd\subseteq \Partial(\Feat)\times\Partial(\Feat)$, %
where $\partial\unlhd\partial'$ stands for $\partial'$ to be \emph{at least as general as}
$\partial$ w.r.t. $\Effects$, is defined by
\[
	\partial\unlhd\partial'\en{ \text{iff} }\Sem{\partial}\cap\Effects \subseteq \Sem{\partial'}\cap\Effects.
\]
The set of \emph{most general causes} $\mCauses$ comprises
those causes for $\Effects$ w.r.t. $\ValidFeat$ that are $\unlhd$-maximal in $\Causes$.
\end{definition}
From the above definition, we directly establish an algorithm to compute
$\mCauses$ by computing $\unlhd$ on $\Causes$ in quadratic time and selecting
elements maximal w.r.t.\ $\unlhd$.
While $\mCauses$ already provides a cause--effect cover of $\Effects$,
there might be different most general causes that cover the same set of effect instances.
Towards nearly minimal cause--effect covers, we thus might pick one of those candidates
with minimal support to obtain an even further concise representation.

\subsection{Responsibility and Blame}\label{sec:responsibility}
To measure the influence of causes on effects, Chockler and
Halpern\ \cite{ChoHal04a} introduced degrees of \emph{responsibility} and \emph{blame},
ranging from zero to one for ``no'' to ``full'' responsibility and blame, respectively.
\emph{Responsibility} measures how relevant a single cause is for an effect in a specific context.
\emph{Blame} denotes the expected overall responsibility according to a given
probability distribution on all contexts where the effect emerges.
We take inspiration from these measures and present corresponding notions
for feature causality. In short, the degree of responsibility is the maximal share
of features contributing to the effect, i.e., features that would have to be
reconfigured to provide a counterfactual witness.

\begin{example}\label{bsp:voteresp}
	We rephrase the \emph{majority example}
	by Chockler and Halpern~\cite{ChoHal04a} in our setting.
	Consider eleven features whose configurations are all
	valid. We are interested in responsibilities for the
	effect that the majority of features is active.
	If all eleven features are active, each feature has
	a responsibility of $1/6$, since six features share the responsibility for the effect:
	Besides the feature of interest, further five features have to be reconfigured
	towards a majority of inactive features. In a configuration where six features are
	active and five not, each of the six active ones is
	fully responsible for the effect: If this feature would be reconfigured,
	more features would be inactive than active.
	We then assign responsibility of one to each of the six active features.
\end{example}

In what follows, we formalize degrees of responsibility and blame for single features 
as in the example above. An extension of this notion to partial configurations
to explicate feature interactions is provided in Section~\ref{sec:featint}.

\tudparagraph{Feature responsibility}
Intuitively, the degree of responsibility of a single feature $x\in\Feat$ is defined as the maximal
share to contribute to causing the effect in an effect instance $\eta\in\Effects$.
In the case that $x$ does not appear in the support of any cause $\gamma$ covering $\eta$,
feature $x$ does not contribute
to causing the effect in $\eta$ and thus has no responsibility.
Otherwise, $x$ shares its responsibility with, at least, a minimal number of
other features, whose switch of its interpretation in $\eta$ would lead to a 
counterfactual witness $\overline{\eta}$, i.e., $\overline{\eta}\in\CEffects$.
We formalize such switches in configurations by a function
\[
	\switch\colon \power{\Feat} \times \Total(\Feat) \ra \Total(\Feat)
\]
where for any $Y\subseteq\Feat$ and $\total\in\Total(\Feat)$, we have
$\switch(Y,\total)(y)=\total(y)$ if $y\not\in Y$ and
$\switch(Y,\total)(y)=\neg\total(y)$ otherwise.

\begin{definition}\label{def:sresp}
	The \emph{degree of responsibility} of a feature $x\in\Feat$
	in the context $\eta\in\Effects$ for which there is $\gamma\in\Causes$ with
	$\eta\in\Sem{\gamma}$ and $x\in\supp(\gamma)$ is defined as
	\[
		\resp(x,\eta) \en{=}
		\frac{1}{
			\min\left\{ |Y| : Y\subseteq\Feat, x\in Y, \switch(Y,\eta)\in\CEffects\right\}
		}
	\]
	and as $\resp(x,\eta)=0$ otherwise.
\end{definition}

Note that due to \ref{en:fc2} there exists at least one counterfactual witness 
in the case $x$ appears in a cause covering $\eta$ and hence, the denominator of 
the above fraction is finite and greater than zero.

\begin{example}\label{bsp:emailresp}
	Continuing the email example (see Example~\ref{bsp:emailcause}),
	the mail feature $m$ and sign feature $s$ do not have any
	responsibility for a long decipher time in any configuration as they do not
	appear in any of the causes $\gamma_a$, $\gamma_r$, and $\gamma_{e\bar{c}}$.
	Also the AES feature $a$ has no responsibility in configurations $mer$
	or $mers$, since the only covering causes are $\gamma_{e\bar{c}}$
	and $\gamma_r$, 	not containing $a$ in their support.
	Besides the analogous case for the RSA feature $r$,
	other degrees of responsibility are $1/2$:
	switching a feature $x$ in $\eta$ usually requires one further
	feature to switch towards a configuration of $\CEffects$.
	For example, selecting the Caesar feature $c$ in $\eta=mea$
	requires also to deselect the AES feature $a$, leading to $\overline{\eta}=mec\in\CEffects$.
\end{example}

\tudparagraph{Blame}
Degrees of responsibility are locally defined w.r.t.\ a context, while one
is surely also interested in a global measure of responsibility of a feature
or partial configuration w.r.t.\ all possible contexts, i.e., effect configurations.
The \emph{degree of blame} is defined as the expected degree of responsibility
on a probability distribution $\pi\colon \ValidFeat \ra [0,1]$
over valid configurations $\ValidFeat$, i.e., where $\sum_{\total\in\ValidFeat}\pi(\total) = 1$.
\begin{definition}\label{def:pblame}
	The \emph{degree of blame} of a feature $x\in\Feat$ w.r.t.\ a distribution $\pi$
	over $\ValidFeat$ is defined as
\[
		\blame(x,\pi) \en{=}
		\sum_{\eta\in\Effects} \pi(\eta)\cdot\resp(x,\eta).
\]
\end{definition}

\begin{example}
	To illustrate the degree of blame, continue Example~\ref{bsp:emailresp}, where
	we assume a uniform distribution $\pi$ over effect configurations, i.e.,
	$\pi(\eta)= 1/|\Effects| =1/4$ for all $\eta\in\Effects$ and $0$ otherwise.
	Then, $\blame(x,\pi) = 1/4\sum_{\eta\in\Effects} \resp(x,\eta)$
	and thus, $\blame(x,\pi)$ is $1/2$ for $x\in\{e,c\}$,
	$3/8$ for $x\in\{a,r\}$, and $0$ for $x\in\{m,s\}$.
\end{example}

\tudparagraph{On the choice of blame distributions}
The distribution $\pi$ models the frequency of valid configurations to occur,
for which there are several scenarios that lead to a reasonable definition of $\pi$.
One natural distribution may model the frequency of users choosing a configuration
in the configurable system. But also the frequency of effect configurations is useful, e.g.,
to model the frequency a certain bug is reported by users when the effect corresponds to the
property of a malfunction. In the case such statistics
are not at hand or one is interested in the degree of blame from a developer's
perspective, uniform distributions over valid configurations or effects
are canonical candidates for $\pi$.

\subsection{Feature Interactions}\label{sec:featint}
A notorious problem in configurable systems is the presence of (inadvertent)
\emph{feature interactions}~\cite{CalKolMag03a,ApeAtlBar+2014}, which describe
system behaviors that emerge due to a combination of multiple features
not easily deducible from the features' individual behaviors.
The detection, isolation, and resolution of feature interactions play a
central role in the development of configurable systems and beyond~\cite{Zave2001,ApeBatKas13a}.
We now show how our black-box causal analysis at the level of features
(see Section~\ref{sec:effects}) can be used for the detection and isolation of
feature interactions. These can provide the basis for fine-grained, white-box 
feature-interaction resolution as done by Garvin and Cohen~\cite{GarCoh11a}.

\tudparagraph{Detection}
The first problem we address is to detect the \emph{necessity} of feature
interactions for an effect to emerge.
Garvin and Cohen presented a formal definition of \emph{feature interaction faults}
to capture faults in configurable systems that
necessarily arise from the interplay between multiple features~\cite{GarCoh11a}.
Notably, their characterization is also at the abstraction level
of features and relies on black-box testing of faults, similar to our perspective
on effects. We transfer their definition to our setting, but covering arbitrary effects
instead of faults only.
Recall that a partial configuration $\omega\in\Partial(\Feat)$ is \emph{sufficient}
for $\Effects$ w.r.t. $\ValidFeat$ if $\varnothing\neq\Sem{\omega}\cap\ValidFeat\subseteq\Effects$
(see \ref{en:fc1}).
\begin{definition}\label{def:fci}
	A partial configuration $\omega\in\Partial(\Feat)$ is a \emph{$t$-way interaction witness}
	for $\Effects$ w.r.t. $\ValidFeat$ if
	\begin{enumerate}[label=\textbf{(FI$_\arabic*$)},leftmargin=*,ref=\textbf{FI$_\arabic*$}]
		\item\label{en:fi1}
			$\omega$ sufficient for $\Effects$ w.r.t. $\ValidFeat$ with $|\supp(\omega)| =t$ and
		\item\label{en:fi2}
			there is no $\hat\omega$ sufficient for $\Effects$ w.r.t. $\ValidFeat$ with $|\supp(\hat\omega)| =t-1$.
	\end{enumerate}

\end{definition}

Basically, there is a one-to-one correspondence between
interaction witnesses and feature causes with minimal support:

\begin{theorem}\label{thm:fis}
	For any partial configuration $\gamma\in\Partial(\Feat)$ with $t=|\supp(\gamma)|$
	we have that\\
	\setlength{\tabcolsep}{1.8pt}
	\begin{tabularx}{\columnwidth}{lX}
          (1)&if $\gamma$ is a $t$-way interaction witness for $\Effects$ w.r.t.\ $\ValidFeat$,
		  then $\gamma\in\Causes$, and\\
	 (2)&if $\gamma\in\Causes$ and there is no $\hat\gamma\in\Causes$ with
                  $|\supp(\hat\gamma)|{<}t$, then $\gamma$ is a $t$-way interaction witness
	          for $\Effects$ w.r.t. $\ValidFeat$.
	\end{tabularx}
\end{theorem}
To this end, Algorithm~\ref{algo:primecause}, in combination with
a projection to feature causes with minimal support, can be used to
decide whether the effect emerges necessarily from feature interactions:
A necessary feature interaction takes place in the case these minimal
feature causes all have a support involving, at least, two features.

\begin{example}\label{bsp:interact}
	Returning to Example~\ref{bsp:emailcause}, there are exactly two causes
	that are both $1$-way interaction witnesses: $\gamma_a$ and $\gamma_r$.
	The cause $\gamma_{e\bar{c}}$ does not witness a necessary $2$-way feature interaction 
	since although having support size of two, $\gamma_a$ and $\gamma_r$ have support size one
    (cf. Theorem~\ref{thm:fis}(2)).
	Hence, the effect describing long decipher time is not
	necessarily related to a feature interaction.
\end{example}

\tudparagraph{Isolation}
The second problem we address is to pinpoint features responsible for
feature interactions. For this, observe that
Definition~\ref{def:fci} is similar to Definition~\ref{def:fc},
but with a different notion of minimality:
While \ref{en:fi2} ensures global minimality over all partial configurations,
\ref{en:fc2} ensures local minimality through expansions, taking
the individual selection of features into account.
To pinpoint those features that actually interact towards the effect,
feature causes can be interpreted as a form of \emph{interaction witnesses} at the
local level instead of the global perspective taken for $t$-way interaction witnesses:
Switching some feature a cause does not ensure the effect to emerge anymore --
hence, the switched feature is necessary for the effect.
For instance, in Example~\ref{bsp:interact}, the interaction between
encryption and Caesar being disabled is witnessed by the
feature cause $\gamma_{e\bar{c}}$.
Feature causes thus also provide a criterion for feature interactions
at the operational level and can be used to guide a more in-depth
white-box feature interaction analysis, possibly reducing the naive 
exponentially-sized feature-interaction search space.

\tudparagraph{Feature interaction responsibility and blame}
Any subset of the supports of feature causes that contains more
than two features provides candidates for an actual interaction
between the features contained.
Since both, the number of feature causes and their expansions, can be
exponential in the number of features, feature interactions isolated
via a causal analysis might still be difficult to interpret by developers.
Based on the degree of responsibility for single features,
we now provide a variant to measure responsibility and blame of
feature interactions, where high values indicate strong relevance
of the interaction and low values weak relevance.

\begin{definition}\label{def:jresp}
	The \emph{degree of responsibility} of a partial configuration $\partial\in\Partial(\Feat)$
	in the context $\eta\in\Effects$ for which there is
	$\gamma\in\Causes$ with $\eta\in\Sem{\gamma}$ and $\supp(\partial)\subseteq\supp(\gamma)$
	and $\partial(x)=\gamma(x)$ for all $x\in\supp(\partial)$ is defined as
	\begin{center}%
	\vspace{-.8em}
	\resizebox{\linewidth}{!}{$\displaystyle\resp(\partial,\eta) \en{=}
		\frac{1}{
			\min\left\{ |Y| : Y\subseteq\Feat, \supp(\partial)\cap Y\neq\varnothing,
				\switch(Y,\eta)\in\CEffects\right\}
		}$}
	\end{center}
	\vspace{.2em}
	and as $\resp(\partial,\eta)=0$ otherwise.
\end{definition}

Note that, for $\partial(x)=\eta(x)$ and $\supp(\partial)=\{x\}$,
Definition~\ref{def:jresp} agrees with Definition~\ref{def:sresp}.
The degree of responsibility is non-zero in the case of a single feature if the feature appears
in some cause, whereas the degree of feature interaction responsibility is non-zero
if some cause is an expansion of the potential feature interaction.
Feature interaction blame is defined as in Definition~\ref{def:pblame} but
with replacing the single feature $x\in\Feat$ by the partial configuration $\partial\in\Partial(\Feat)$
standing for the potential feature interaction of interest.

\begin{table*}[t]
	\caption{\label{tab:stats}Statistics of feature causality experiments}
	\Description{Statistics of feature causality experiments}
	\footnotesize
	\resizebox{\textwidth}{!}{%

\begin{tabular}{cl|rrrr|rrrr}
\toprule
Property\!\!\!&\multicolumn{1}{l}{System}&\#&$|\ValidFeat|$&$|\Feat|$&\multicolumn{1}{r}{Time}&\multicolumn{4}{c}{Average size of}\\
\cmidrule(lr){7-10}
{\tiny[type]}&\multicolumn{1}{c}{}& &&&\multicolumn{1}{r}{\tiny[s]}&   $\Effects$ & $\cC$ & $m\cC$ &   DLS \\
\midrule
\multirow{3}{*}{$\varphi$}&\textsc{CFDP} & 10 &  $           56$ &  13 &    {0.21} &      {28} &    {7.30} &     {2.60} & {58\%} \\
	&\elevator$\!_1$ & 36 &  $          256$ &   9 &        {0.09} &     {128} &    {0.58} &     {0.58} & {26\%} \\
   	&\minepump & 82 &  $          128$ &  11 &        {0.29} &      {64} &    {1.77} &     {0.91} & {43\%} \\\midrule
\multirow{6}{*}{$\varepsilon_B$}
	&\textsc{LinkedList} & 42 &  $          204$ &  19 &                  {18.97} &     {102} &   {47.40} &     {8.36} & {54\%} \\
	&\linux & 21 &  $ 16\,777\,216$ &  25 &                   {5.88} &      {22} &   {22.14} &    {22.14} & {80\%} \\
	&\textsc{PKJab}& 28 &  $           72$ &  12 &            {0.42} &      {36} &    {8.43} &     {4.86} & {72\%} \\
	&\textsc{Prevaylar}& 42 &  $           24$ &   7 &        {0.20} &      {12} &    {3.57} &     {2.17} & {93\%} \\
	&\textsc{SNW} & 42 &  $       3\,240$ &  36 &          {2\,751.32} &    {1\,620} & {2\,110.31} &    {18.90} & {37\%} \\
        & \textsc{ZipMe}~$\!_1$ & 42 &  $          64$ &  9 &                  {0.19} &     {32} &    {4.19} &     {2.69} & {93\%}\\\midrule
\multirow{4}{*}{$\varepsilon_M$}
	&\textsc{Curl}& 28 &  $          768$ &  14 &                  {83.07} &     {384} &   {96.75} &    {67.39} & {59\%} \\
        & \textsc{h264}& 28 &  $       1\,152$ &  17 &                {355.37} &     {576} &  {229.86} &    {113.93} & {53\%} \\
	&\textsc{SQLite} & 21 &  $  3\,932\,160$ &  40 &             {34\,826.86} & {1\,310\,729} & {1\,881.67} &   {436.48} & {27\%} \\
	&\textsc{WGet} & 28 &  $       5\,120$ &  17 &                 {777.26} &    {2\,560} &  {298.11} &   {201.79} & {51\%} \\
\bottomrule
\end{tabular}
\hspace{1em}
\begin{tabular}{cl|rrrr|rrrr}
\toprule
Property\!\!\!&\multicolumn{1}{l}{System}&\#&$|\ValidFeat|$&$|\Feat|$&\multicolumn{1}{r}{Time}&\multicolumn{4}{c}{Average size of}\\
\cmidrule(lr){7-10}
{\tiny[type]}&\multicolumn{1}{c}{}& &&&\multicolumn{1}{r}{\tiny[s]}&   $\Effects$ & $\cC$ & $m\cC$ &  DLS \\
\midrule
\multirow{5}{*}{$\varepsilon_T$}
	&\apache & 28 &  $          192$ &  10 &                   {2.85} &      {96} &   {28.79} &    {22.96} & {69\%} \\
	&\elevator$\!_2$ & 28 &  $           10$ &   6 &                   {0.14} &       {5} &    {3.21} &     {2.25} & {80\%} \\
	&\Email & 28 &  $           40$ &  10 &                   {1.01} &      {20} &   {22.32} &     {8.21} & {69\%} \\
        & \textsc{h264}& 56 &  $       1\,152$ &  17 &                {32.24} &     {576} &  {68.98} &    {22.71} & {51\%} \\
        & \textsc{ZipMe}~$\!_2$& 42 &  $          640$ &  16 &                  {1.73} &     {320} &    {11.07} &     {9.98} & {72\%}\\\midrule
\multirow{2}{*}{$\thresh_R$}
	&\textsc{BSN} &  4 &  $          298$ &  11 &     {0.63} &     {149} &   {21.00} &     {8.75} & {68\%} \\
    &\textsc{VCL} & 22 &  $    2\,097\,152$ &  21 &    {60\,403.78} & {1\,048\,576} & {3\,718.05} &  {3\,718.05} & {36\%} \\\midrule
\multirow{5}{*}{$\thresh_T$}
	&\berkeley & 36 &  $       2\,560$ &  19 &       {5.71} &    {1\,280} &   {13.03} &     {4.28} & {72\%} \\
	&\dune & 46 &  $       2\,304$ &  32 &    {3\,950.15} &    {1\,152} &  {724.87} &    {46.00} & {46\%} \\
	&\llvm & 30 &  $       1\,024$ &  12 &      {16.21} &     {512} &   {53.57} &    {53.57} & {55\%} \\
	&\lrzip & 48 &  $          432$ &  20 &       {5.71} &     {216} &   {13.85} &     {1.90} & {69\%} \\
        &\xtwosixfour & 48 &  $       1\,152$ &  17 &       {2.38} &     {576} &    {7.50} &     {3.75} & {78\%} \\[1.14em]
\bottomrule
\end{tabular}
 }
        ~\\[.5em]\footnotesize
        We list for each type of effect property (functional LTL properties $\varphi$, 
        precision accuracy threshold properties $\varepsilon$,
        and other threshold properties $\thresh$)
        the considered subject systems, the numbers of experiments (\#), 
        valid configurations ($|\ValidFeat|$), features ($|\Feat|$), and the overall time in seconds to
		compute feature causes and most general causes.
        The second part of the table lists the average sizes of the effect sets ($\Effects$), feature causes
        ($\cC=\Causes$), cause--effect covers by most general causes ($m\cC=\mCauses$), and
        DLS formulas relative to the characteristic formula of causes 
        (DLS=$|\reduce(\chi(\cC))|/|\chi(\cC)|$
		in percent).
\end{table*}

\section{Experiment Setup}\label{sec:casestudy}
To evaluate our causal analysis and explication algorithms, we conducted
a number of experiments comprising many analyses on community benchmarks
and real-world examples from the area of configurable software systems.

\subsection{Research Questions}
Our evaluation is driven by four research questions that address the key issue of whether
and how the notion of feature causality facilitates identifying root causes, estimating
the effects of features, and detecting feature interactions in controlled and practical settings.

\begin{enumerate}[label=\textbf{(RQ$_\arabic*$)},leftmargin=*,ref=\textbf{RQ$_\arabic*$}]
	\item\label{rq:effective}
		Can feature causes be effectively computed in real-world settings and
		support the detection of reasons for different effects of interest?
	\item\label{rq:explicate}
		Do DLS representation, cause--effect covers, and responsibility
		and blame degrees provide concise causal explications?
	\item\label{rq:guide}
		Is feature causality beneficial for guiding the configuration of
		systems under variability-aware constraints?
	\item\label{rq:interaction}
		Can feature interactions and configuration-dependent anomalies
		be detected and isolated based on feature causality?
\end{enumerate}

\subsection{Implementation}
We implemented our algorithms to compute feature causes and explications in the prototypical
tool \featcause{}. Written in Python, our tool relies on the engines for logical expressions
and binary decision diagrams (BDDs) of \pyeda{}, a library for electronic design
automation~\cite{pyeda}.
The tool takes the sets of valid feature configurations $\ValidFeat$ and effects $\Effects$
as input. \featcause{} supports different input formats for these sets,
e.g., by Boolean expressions in DNF or CNF.
Internally, sets of (partial) feature configurations are efficiently represented as
reduced ordered BDDs~\cite{Bryant86}.
In addition to their compact and hence space-efficient representation, we chose BDDs because they
provide an efficient method to check satisfiability (required, e.g., for Line~\ref{l:remove} in
Algorithm~\ref{algo:primecause}).
We first implemented a naive algorithm directly checking the conditions \ref{en:fc1} and \ref{en:fc2},
but even for small examples, we easily ran into timeouts.
Hence, we devised Algorithm~\ref{algo:primecause}, which
uses prime implicants to efficiently determine feature causes (cf. Section~\ref{sec:causality}).
To compute prime implicants, we used the tool \espresso{}~\cite{McGSanBra93a}, well known from 
circuit optimization, through an interface that mediates between
our BDD representations and the DNF representations in \espresso{}'s PLA format.
This interface is also used to provide minimal and nearly minimal cause--effect covers through
\signature{} and \espresso{}, respectively, which can then
be compared with our heuristic cause--effect cover by most general causes (see Section~\ref{sec:smin}).
While it is well known that the length of DNFs can be exponential in the size of the BDD
representing the same Boolean function, generating DNFs from our BDD representations
did not face any significant blowup and required negligible time in all our experiments.
Besides the core tool, we have implemented several conversion scripts to generate valid
feature configuration sets from \tvl{}~\cite{tvl2011} and effect sets from analysis results
returned by variability-aware analysis tools such as \provelines{}~\cite{ProveLines13} and 
\profeat{}~\cite{CDKB18} (see also Section~\ref{sec:functional})
and the data sets from Siegmund et al.~\cite{SieGreApe15a} and Kaltenecker et al.~\cite{KalGreSie2019}.

\subsection{Subject Systems}
To answer our research questions, we selected a diverse set
of subject systems, ranging from popular community benchmarks
to more involved systems with non-functional properties and from
real-world settings (see Table~\ref{tab:stats} for an overview).

From Cordy et al.~\cite{ProveLines13}, we use \minepump{}, \elevator{}, and \cfdp systems and
analyzed them against the accompanied LTL properties using the variability-aware
model checker \provelines{}. Furthermore, we took the \Email{} and \elevator{}
from von Rhein et al.~\cite{RheGreApe15a} and analyzed multiple defects provided as propositional logic formula
generated by \splverifier{} \cite{ApeRheWen2013}.

For quantitative properties, we generated effect sets
from configurable system analysis results as illustrated in Section \ref{sec:nonfunctional}
for three classes of systems. First, we considered configurable systems modeled
for the variability-aware probabilistic model checker \profeat{}~\cite{CDKB18},
comprising a body sensor network (\textsc{BSN}) model \cite{RodANLCSSL15}
and a velocity control loop (\textsc{VCL}) model of an aircraft \cite{DDMBJ19,DubMorBai20b}.
In both systems, the reliability (R) of the system is analyzed in terms of the probability
of failure of sensors and control components, respectively. Second, we generated
effect sets from performance measurements of real-world configurable software systems that
have been used to evaluate performance modeling techniques~\cite{SieGreApe15a,KalGreSie2019}.
In particular, we selected five systems from different domains: a compiler framework (\llvm),
a database-system (\berkeley), a compression tool (\lrzip), a video encoder (\xtwosixfour), and a
toolbox for solving partial differential equations (\dune).
For these systems, we chose thresholds on the runtime (T) of the system executions.
Third, we constructed effect sets for several systems from studies on performance prediction of non-functional properties~\cite{SieRosKaeGia2013,SieKolKaeApe2012,SieRheApe2012},
such as \apache, \linux, \SQLite, and \wget.
For these, we used the black-box approach by Siegmund et al.~\cite{SieRosKuh+12},
which uses multivariable linear regression methods to generate variability-aware performance
models. Our thresholds for constructing effect sets are imposed on the prediction accuracy
of the following three different non-functional properties on those systems:
runtime (T), binary size (B), and memory footprint (M).

\subsection{Operationalization}
For ~\ref{rq:effective}, we compute both sets $\Causes$ and $\CCauses$, i.e.,
the feature causes of the effect property and its negation.
Based on these feature causes, we further compute cause--effect covers $\mCauses$
and $\mCCauses$, distributive law simplification (DLS),
as well as responsibility and blame values for each single feature and cause to answer \ref{rq:explicate}.
For blame computations, we assume a uniform distribution over all effects,
due to the absence of further statistical information and taking a
developers' perspective (cf. Section~\ref{sec:responsibility}).
For~\ref{rq:guide}, we compute single feature blames based on a uniform
effect distribution to measure the influence of individual features onto the effect.
We compute feature interaction blames on pairs of features to address~\ref{rq:interaction},
again assuming a uniform effect distribution.
Table~\ref{tab:stats} provides key statistics about our experiments, focusing
on model characteristics and the time to compute feature causes and most general
feature causes. All experiments were conducted on an AMD Ryzen~7 3800X 8-Core system
with 32GB of RAM\ running Debian~10 and Python 3.7.3.

\section{Results}\label{sec:results}
We discuss our results w.r.t.\ the different kinds of causal explications
of Section~\ref{sec:minimal}. First, we discuss statistics on our experiments,
also quantitatively analyzing the potential of causal explications by
most general causes and DLS-reduced formulas.
Then, we address our research questions in more depth by means of three representative
subject systems. Here, we focus on properties not detectable by classical
causal white-box analysis methods~\cite{Zel02a,JohBruMel20}.

\subsection{Descriptive Statistics (\ref{rq:effective} and \ref{rq:explicate})}
The computability of feature causes is of major interest for our evaluation. Table~\ref{tab:stats} provides
an overview of the subject systems for which we generated effect sets and applied our feature causality analysis.
We see that our algorithms compute feature causes in reasonable time, within a few seconds for most subject
systems. To create the effect sets, we considered several effect properties, as described in Section~\ref{sec:effects}:
$\varphi$ stands for LTL properties; 
$\varepsilon_B$, $\varepsilon_M$, and $\varepsilon_T$ for thresholds on the accuracy of a prediction
model for \underline{b}inary size, \underline{m}emory footprint, and run\underline{t}ime;
and $\thresh_R$ and $\thresh_T$ for
\underline{r}eliability and run\underline{t}ime  thresholds, respectively. 
This variety of properties already illustrates
the wide range of applications and potential use of feature causality.
The sizes of valid configuration and effect sets crucially influence the time for
computing feature causes, which is as expected since the complexity of
Algorithm~\ref{algo:primecause} is dominated by the computation of prime implicants
of $\big(\Total(\Feat){\setminus}\ValidFeat\big)\cup\Effects$. 
Since our implementation relies on BDDs for the representation of valid configuration
and effect sets, it is however well possible that within similar sizes,
computation times can significantly differ. 
This is mainly due to the fact that BDD sizes highly depend on the specific nature of the
represented Boolean functions and the variable order chosen \cite{Bryant86}.
For instance, while the experiments on \textsc{DUNE}
and \textsc{BerkeleyDB} (see Table~\ref{tab:stats}) have similar sized sets $\ValidFeat$
and $\Effects$, their runtimes differ in two orders of magnitude.
We see that the number of most general feature causes is often way smaller than the 
overall number of feature causes, which renders the creation of cause--effect covers
by most general causes sensible to support concise explications. 
Interestingly, cause--effect covers by most general causes and \espresso\
minimization~\cite{McGSanBra93a} yield the same results in almost all of our experiments,
which proves our heuristics to be effective.
In the same vein, the application of DLS
leads to great reductions of logical representations of feature causes, e.g., 
on average by almost 3/4 in the \elevator$\!_1$ subject system (see Table~\ref{tab:stats}).\\

\begin{mdframed}[backgroundcolor=lightgray!30,%
	linecolor=lightgray!30,%
	innertopmargin = 3pt,%
	innerleftmargin = 3pt,%
	innerrightmargin = 3pt]
For~\ref{rq:effective} and~\ref{rq:explicate}, we conclude that feature causes
are computable in reasonable time. A substantial reduction of the set of feature causes
and cause--effect covers can be performed with
DLS formulas and most general feature causes, respectively.
\end{mdframed}

\begin{table}[t]
	\caption{\label{tab:minepump}\minepump\ -- (most general) feature causes}
	\Description{Feature causes for minepump}
	\resizebox{\columnwidth}{!}{
	\begin{tabular}{l|l}\toprule
		Blame &
		Feature cause (characteristic formula)\\\midrule
		0.29 &
		$\mathrm{High}\wedge\mathrm{Command}\wedge\neg\mathrm{Stop}\wedge\mathrm{MethaneAlarm}$ \\
		\rowcolor{blue!10}
		0.57 &
		$\mathrm{High}\wedge\mathrm{Start}\wedge\mathrm{MethaneAlarm}$\\
		0.14 &
		$\mathrm{High}\wedge\mathrm{Command}\wedge\neg\mathrm{Stop}\wedge\mathrm{MethaneSensor}\wedge\neg\mathrm{MethaneQuery}$\\
		\rowcolor{blue!10}
		0.57 &
		$\mathrm{High}\wedge\mathrm{Start}\wedge\mathrm{Stop}$\\
		\rowcolor{blue!10}
		0.57 &
		$\mathrm{High}\wedge\mathrm{Low}\wedge\mathrm{Start}$\\
		0.29 &
		$\mathrm{High}\wedge\mathrm{Start}\wedge\mathrm{MethaneSensor}\wedge\neg\mathrm{MethaneQuery}$\\
		0.29 &
		$\mathrm{High}\wedge\mathrm{Low}\wedge\mathrm{Command}\wedge\neg\mathrm{Stop}$\\\bottomrule
	\end{tabular}}
\end{table}

\subsection{Feature Cause Explications (\ref{rq:explicate})}
We discuss the explications of feature causes that we have generated by the example of
the \minepump\ system~\cite{ClaCorSchobHeyLegRas13}, which is frequently used in the
configurable systems' analysis community.
This system models a water pump of a mine with $|\Feat| = 11$ features on which
requirements expressed in LTL are imposed (see Section~\ref{sec:effects}).
An analysis of the stabilization property, formalizing that the
\minepump system eventually stabilizes (the pumps are eventually continuously on or off),
using \provelines{} returned $|\Effects| =28$ configurations where the property holds
and $|\ValidFeat{\setminus}\Effects| =100$ configurations where it does not hold.

The direct interpretation of features responsible for this effect property is
difficult, as it requires to investigate the result of all $|\ValidFeat| = 128$
configurations. A causal analysis returned seven feature causes listed %
with their degree of blame in Table~\ref{tab:minepump}. They already provide hints
which features are responsible for the property. Among the feature causes, three are most general,
highlighted in Table~\ref{tab:minepump}. They have the highest degree of partial configuration
blame while the most lengthy cause has the smallest degree. Our DLS
heuristics on most general causes yields %
\[
	\mathrm{High} \wedge \mathrm{Start} \wedge
	(\mathrm{Stop} \vee \mathrm{Low} \vee \mathrm{MethaneAlarm})
\]
providing a concise representation of feature cause candidates
explicating the effect property.
That is, selecting features $\mathrm{High}$, $\mathrm{Start}$,
and one of the three features $\mathrm{Stop}$, $\mathrm{Low}$, or $\mathrm{MethaneAlarm}$
covers all causally relevant configurations for
the \minepump\ system to stabilize.
On other subject systems, explications are also effective, but
with less drastic reductions than for the \minepump\ example.\\[-.4em]
\begin{mdframed}[backgroundcolor=lightgray!30,%
	linecolor=lightgray!30,%
	innertopmargin = 3pt,%
	innerleftmargin = 3pt,%
	innerrightmargin = 3pt]
Answering~\ref{rq:explicate}, feature causes are of reasonable size
compared to the complete analysis results, i.e., most general feature causes 
and DLS provide concise explications for feature causes.
Responsibility and blame reflect the impact of feature causes.
\end{mdframed}
\begin{table*}[htbp]
	\parbox{.6\textwidth}{
	\centering\small
	\caption{\label{tab:redundancy}Feature blame for the \textsc{VCL} redundancy system}
	\resizebox{.6\textwidth}{!}{
	  \begin{tabular}{rccccccccccccc}
		\multicolumn{1}{c}{\parbox{.4cm}{\centering $\thresh_R$ {\tiny [$10^{-2}$]}}}	& \RotText{MFunc} & \multicolumn{1}{p{2.5em}}{\RotText{Sum1, Sum4}} & \RotText{Memory} & \multicolumn{1}{p{2.5em}}{\RotText{MOne, P, vCruise}} & \RotText{SpentFuel, I} & \RotText{D} & \RotText{Prod} & \multicolumn{1}{p{2.5em}}{\RotText{DragForce, EThrust}} & \RotText{Integrator} & \RotText{Sum5} & \RotText{Sum3} & \RotText{Sum2} & \RotText{Acceleration} \\\midrule
	  1.9 & \cellcolor[rgb]{ .784,  .827,  .427}0.63 & \cellcolor[rgb]{ .439,  .678,  .278}1.00 & \cellcolor[rgb]{ .784,  .827,  .427}0.63 & \cellcolor[rgb]{ .439,  .678,  .278}1.00 & \cellcolor[rgb]{ .439,  .678,  .278}1.00 & \cellcolor[rgb]{ .439,  .678,  .278}1.00 & \cellcolor[rgb]{ .784,  .827,  .427}0.63 & \cellcolor[rgb]{ .439,  .678,  .278}1.00 & \cellcolor[rgb]{ .439,  .678,  .278}1.00 & \cellcolor[rgb]{ .439,  .678,  .278}1.00 & \cellcolor[rgb]{ .439,  .678,  .278}1.00 & \cellcolor[rgb]{ .439,  .678,  .278}1.00 & \cellcolor[rgb]{ .439,  .678,  .278}1.00 \\
	  2.0 & \cellcolor[rgb]{ .714,  .796,  .396}0.70 & \cellcolor[rgb]{ .624,  .757,  .357}0.80 & \cellcolor[rgb]{ .647,  .769,  .369}0.77 & \cellcolor[rgb]{ .608,  .753,  .353}0.82 & \cellcolor[rgb]{ .494,  .702,  .302}0.94 & \cellcolor[rgb]{ .439,  .678,  .278}1.00 & \cellcolor[rgb]{ .714,  .796,  .396}0.70 & \cellcolor[rgb]{ .608,  .753,  .353}0.82 & \cellcolor[rgb]{ .439,  .678,  .278}1.00 & \cellcolor[rgb]{ .624,  .757,  .357}0.80 & \cellcolor[rgb]{ .624,  .757,  .357}0.80 & \cellcolor[rgb]{ .624,  .757,  .357}0.80 & \cellcolor[rgb]{ .439,  .678,  .278}1.00 \\
	  2.2 & \cellcolor[rgb]{ .902,  .882,  .478}0.49 & \cellcolor[rgb]{ .851,  .859,  .455}0.55 & \cellcolor[rgb]{ .851,  .859,  .455}0.55 & \cellcolor[rgb]{ .796,  .835,  .431}0.61 & \cellcolor[rgb]{ .671,  .78,  .376}0.75 & \cellcolor[rgb]{ .49,  .702,  .302}0.95 & \cellcolor[rgb]{ .902,  .882,  .478}0.49 & \cellcolor[rgb]{ .796,  .835,  .431}0.61 & \cellcolor[rgb]{ .439,  .678,  .278}1.00 & \cellcolor[rgb]{ .851,  .859,  .455}0.55 & \cellcolor[rgb]{ .851,  .859,  .455}0.55 & \cellcolor[rgb]{ .851,  .859,  .455}0.55 & \cellcolor[rgb]{ .439,  .678,  .278}1.00 \\
	  2.5 & \cellcolor[rgb]{ 1,  .941,  .639}0.35 & \cellcolor[rgb]{ 1,  .929,  .545}0.38 & \cellcolor[rgb]{ 1,  .937,  .612}0.36 & \cellcolor[rgb]{ .988,  .918,  .514}0.40 & \cellcolor[rgb]{ .937,  .894,  .494}0.46 & \cellcolor[rgb]{ .761,  .82,  .416}0.65 & \cellcolor[rgb]{ 1,  .941,  .639}0.35 & \cellcolor[rgb]{ .988,  .918,  .514}0.40 & \cellcolor[rgb]{ .443,  .682,  .282}1.00 & \cellcolor[rgb]{ 1,  .929,  .545}0.38 & \cellcolor[rgb]{ 1,  .929,  .545}0.38 & \cellcolor[rgb]{ 1,  .929,  .545}0.38 & \cellcolor[rgb]{ .439,  .678,  .278}1.00 \\
	  2.9 & \cellcolor[rgb]{ 1,  .976,  .855}0.28 & \cellcolor[rgb]{ 1,  .973,  .831}0.29 & \cellcolor[rgb]{ 1,  .976,  .839}0.28 & \cellcolor[rgb]{ 1,  .969,  .808}0.29 & \cellcolor[rgb]{ 1,  .965,  .765}0.31 & \cellcolor[rgb]{ 1,  .949,  .678}0.33 & \cellcolor[rgb]{ 1,  .976,  .855}0.28 & \cellcolor[rgb]{ 1,  .969,  .808}0.29 & \cellcolor[rgb]{ .522,  .718,  .314}0.91 & \cellcolor[rgb]{ 1,  .973,  .831}0.29 & \cellcolor[rgb]{ 1,  .973,  .831}0.29 & \cellcolor[rgb]{ 1,  .973,  .831}0.29 & \cellcolor[rgb]{ .443,  .682,  .282}1.00 \\
	  3.4 & \cellcolor[rgb]{ 1,  .976,  .851}0.28 & \cellcolor[rgb]{ 1,  .973,  .816}0.29 & \cellcolor[rgb]{ 1,  .969,  .796}0.30 & \cellcolor[rgb]{ 1,  .969,  .788}0.30 & \cellcolor[rgb]{ 1,  .957,  .725}0.32 & \cellcolor[rgb]{ 1,  .925,  .533}0.38 & \cellcolor[rgb]{ 1,  .976,  .851}0.28 & \cellcolor[rgb]{ 1,  .969,  .788}0.30 & \cellcolor[rgb]{ .835,  .851,  .447}0.57 & \cellcolor[rgb]{ 1,  .973,  .816}0.29 & \cellcolor[rgb]{ 1,  .965,  .78}0.30 & \cellcolor[rgb]{ 1,  .965,  .78}0.30 & \cellcolor[rgb]{ .467,  .69,  .29}0.97 \\
	  4.0 & \cellcolor[rgb]{ 1,  .988,  .918}0.26 & \cellcolor[rgb]{ 1,  .984,  .894}0.27 & \cellcolor[rgb]{ 1,  .984,  .89}0.27 & \cellcolor[rgb]{ 1,  .98,  .878}0.27 & \cellcolor[rgb]{ 1,  .976,  .851}0.28 & \cellcolor[rgb]{ 1,  .965,  .769}0.31 & \cellcolor[rgb]{ 1,  .988,  .918}0.26 & \cellcolor[rgb]{ 1,  .98,  .878}0.27 & \cellcolor[rgb]{ 1,  .929,  .565}0.37 & \cellcolor[rgb]{ 1,  .984,  .894}0.27 & \cellcolor[rgb]{ 1,  .949,  .667}0.34 & \cellcolor[rgb]{ 1,  .949,  .667}0.34 & \cellcolor[rgb]{ .573,  .737,  .337}0.86 \\
	  4.7 & \cellcolor[rgb]{ 1,  .996,  .961}0.24 & \cellcolor[rgb]{ 1,  .992,  .953}0.25 & \cellcolor[rgb]{ 1,  .992,  .953}0.25 & \cellcolor[rgb]{ 1,  .992,  .949}0.25 & \cellcolor[rgb]{ 1,  .992,  .937}0.25 & \cellcolor[rgb]{ 1,  .988,  .906}0.26 & \cellcolor[rgb]{ 1,  .996,  .961}0.25 & \cellcolor[rgb]{ 1,  .992,  .949}0.25 & \cellcolor[rgb]{ 1,  .965,  .78}0.30 & \cellcolor[rgb]{ 1,  .992,  .953}0.25 & \cellcolor[rgb]{ 1,  .937,  .604}0.36 & \cellcolor[rgb]{ 1,  .937,  .6}0.36 & \cellcolor[rgb]{ .639,  .765,  .365}0.78 \\
	  5.5 & \cellcolor[rgb]{ 1,  1,  .98}0.24 & \cellcolor[rgb]{ 1,  .996,  .973}0.24 & \cellcolor[rgb]{ 1,  .996,  .973}0.24 & \cellcolor[rgb]{ 1,  .996,  .969}0.24 & \cellcolor[rgb]{ 1,  .996,  .957}0.25 & \cellcolor[rgb]{ 1,  .988,  .918}0.26 & \cellcolor[rgb]{ 1,  .996,  .961}0.24 & \cellcolor[rgb]{ 1,  .992,  .949}0.25 & \cellcolor[rgb]{ 1,  .98,  .867}0.27 & \cellcolor[rgb]{ 1,  .988,  .918}0.26 & \cellcolor[rgb]{ 1,  .933,  .573}0.37 & \cellcolor[rgb]{ .992,  .922,  .518}0.39 & \cellcolor[rgb]{ .729,  .804,  .404}0.68 \\
	  6.4 & 0.23  & 0.23  & 0.23  & \cellcolor[rgb]{ 1,  1,  .992}0.23 & \cellcolor[rgb]{ 1,  1,  .988}0.24 & \cellcolor[rgb]{ 1,  .996,  .973}0.24 & \cellcolor[rgb]{ 1,  .996,  .965}0.24 & \cellcolor[rgb]{ 1,  .992,  .953}0.25 & \cellcolor[rgb]{ 1,  .984,  .882}0.27 & \cellcolor[rgb]{ 1,  .98,  .867}0.27 & \cellcolor[rgb]{ 1,  .941,  .627}0.35 & \cellcolor[rgb]{ .976,  .914,  .51}0.41 & \cellcolor[rgb]{ .792,  .831,  .431}0.61 \\
   \end{tabular}%
	}}
	\hfill
	\parbox{.38\textwidth}{\centering\small
	\caption{\label{tab:lrzip}Feature interaction blame for \lrzip}
	\vspace{-.35cm}
	  \resizebox{.38\textwidth}{!}{
	  \begin{tabular}{c|cccccccc}
		\multicolumn{1}{c}{$\thresh_T$} & \multicolumn{2}{c}{Gzip} & \multicolumn{4}{c}{Lrzip} & \multicolumn{2}{c}{Zpaq} \\\cmidrule{2-3} \cmidrule(lr{.75em}){4-7} \cmidrule{8-9}
		\multicolumn{1}{c}{\tiny [$10^2$s]} & $8$ & $9$ & $4$--$6$ & $7$ & $8$ & $9$ & $4$--$7$ & $8$--$9$ \\\midrule
	  2     &     0.002 & 0.009 &       &       &       &       &       &  \\
	  3     &           &       & \cellcolor[rgb]{ 1,  .996,  .965}0.033 & \cellcolor[rgb]{ 1,  .996,  .965}0.033 & \cellcolor[rgb]{ 1,  .996,  .965}0.033 & \cellcolor[rgb]{ 1,  .996,  .965}0.033 &       &  \\
	  4     &           &       &       & \cellcolor[rgb]{ 1,  .98,  .871}0.042 & \cellcolor[rgb]{ 1,  .98,  .871}0.042 & \cellcolor[rgb]{ 1,  .98,  .871}0.042 &       &  \\
	  5     &           &       &       & \cellcolor[rgb]{ 1,  .98,  .871}0.042 & \cellcolor[rgb]{ 1,  .98,  .871}0.042 & \cellcolor[rgb]{ 1,  .98,  .871}0.042 &       &  \\
	  6     &           & \textcolor{gray!50}{1-way} &       & \cellcolor[rgb]{ 1,  .98,  .871}0.042 & \cellcolor[rgb]{ 1,  .98,  .871}0.042 & \cellcolor[rgb]{ 1,  .98,  .871}0.042 &       &  \\\hline
	  7     &           & \textcolor{gray!50}{2-way} &       & \cellcolor[rgb]{ 1,  .957,  .714}0.056 & \cellcolor[rgb]{ 1,  .957,  .714}0.056 & \cellcolor[rgb]{ 1,  .957,  .714}0.056 & \cellcolor[rgb]{ 1,  .957,  .714}0.056 & \cellcolor[rgb]{ 1,  .957,  .714}0.056 \\
	  8     &           &       &       & \cellcolor[rgb]{ 1,  .957,  .714}0.056 & \cellcolor[rgb]{ 1,  .957,  .714}0.056 & \cellcolor[rgb]{ 1,  .957,  .714}0.056 & \cellcolor[rgb]{ 1,  .957,  .714}0.056 & \cellcolor[rgb]{ 1,  .957,  .714}0.056 \\
	  9     &           &       &       & \cellcolor[rgb]{ 1,  .941,  .635}0.063 & \cellcolor[rgb]{ 1,  .941,  .635}0.063 &       & \cellcolor[rgb]{ 1,  .941,  .635}0.063 & \cellcolor[rgb]{ 1,  .941,  .635}0.063 \\
	  10    &           &       &       & \cellcolor[rgb]{ 1,  .941,  .635}0.063 & \cellcolor[rgb]{ 1,  .941,  .635}0.063 &       & \cellcolor[rgb]{ 1,  .941,  .635}0.063 & \cellcolor[rgb]{ 1,  .941,  .635}0.063 \\
	  11    &           &       &       &       & \cellcolor[rgb]{ 1,  .925,  .533}0.071 &       & \cellcolor[rgb]{ 1,  .925,  .533}0.071 & \cellcolor[rgb]{ 1,  .925,  .533}0.071 \\
	  12    &           &       &       &       &       &       & \cellcolor[rgb]{ .737,  .859,  .392}0.083 & \cellcolor[rgb]{ .737,  .859,  .392}0.083 \\
	  $\cdots$&           &       &       &       &       &       & \cellcolor[rgb]{ .737,  .859,  .392} & \cellcolor[rgb]{ .737,  .859,  .392} \\[-.15cm]
	  22    &      		&       &       &       &       &       & \cellcolor[rgb]{ .737,  .859,  .392}0.083 & \cellcolor[rgb]{ .737,  .859,  .392}0.083 \\
	  23    &           &       &       &       &       &       &       & \cellcolor[rgb]{ .439,  .678,  .278}0.250 \\
	  \end{tabular}%
	}}
\end{table*}%
\subsection{Causality-guided Configuration (\ref{rq:guide})}\label{subsec:AVCL}
Feature blame provides a quantitative measure on the causal impact of feature selections 
w.r.t. a set of configurations. This measure can be used to support configuration decisions, 
e.g., by prioritizing features with high blame values in case the effect property is desirable.
We investigate such a causality-guided configuration on
the velocity control loop (\textsc{VCL}) subject system~\cite{DubMorBai20a,DubMorBai20b}.
The \textsc{VCL} models an aircraft velocity controller in \simulink{} for which
its reliability in terms of probability of failure is of interest.
A common principle to increase the reliability of a system is by \emph{triple modular
redundancy (TMR)} where system components are triplicated and their output are combined via
a majority vote. Dubslaff et al.~\cite{DDMBJ19} suggested to model and analyze systems
with such \emph{protection mechanisms} using family-based methods from configurable systems' 
analysis. To each component they assign a \emph{protection feature} that specifies whether a
component is triplicated or not.
Comprising 21 components eligible for protection, %
the \textsc{VCL} model has $|\ValidFeat| =2^{21}=2\,097\,152$ valid feature configurations.
Clearly, the highest reliability is achieved by protecting all components.
However, each protection comes at its costs in terms of execution time, energy costs, and
packaging size.
While it is known how to determine protection configurations with
optimal reliability--cost tradeoff~\cite{DDMBJ19}, reasons for why a protection configuration
is optimal or why a component was selected for protection are typically unclear.
We address this issue exploiting our causal analysis methods.
Using the variability-aware probabilistic model checking tool \profeat{}~\cite{CDKB18},
we generated effect sets $\Effects_{\rho<\thresh_R}$ w.r.t.\ $\rho$ mapping to
the probability of failure of the \textsc{VCL} within two 
control-loop executions and reliability thresholds $\thresh_R$ between $0.019$ and 
$0.064$\footnote{
	To ensure timely analysis results, real-world failure probability measures
	were increased by a factor $100$~\cite{DubMorBai20a}.
	Resulting higher values might seem unrealistic
	but arguably do not affect the causal analysis measuring the
	impact of protections.
}.
Table~\ref{tab:redundancy} shows the degree of feature blame for 18 protection features of
the 21 components of VCL %
(cf. Section~\ref{sec:responsibility}).
The three components not shown in the table are input components, having zero degree of
blame and hence do not contribute to the systems' reliability.
With the lowest threshold $\thresh_R=0.019$,
the effect set contains only 32 out of the $2^{21}$ feature configurations;
almost all protections of components are responsible for the low probability of failure.
Increasing the threshold lowers the degree of blame since the effect set increases,
leading to less counterfactual witnesses. Using our blame analysis,
we can directly deduce advice for engineers about components protections towards high
reliability: With tight reliability constraints, one should protect the ``Acceleration''
component, followed by the ``Integrator'' component, as their blames are significantly
higher than for other components (cf. upper rows of Table~\ref{tab:redundancy}).
When higher failure rates are acceptable, one should prefer to protect the components
``Sum2'' and ``Sum3'' instead of the ``Integrator'' component due to their higher
impact on reliability (cf. lower rows of Table~\ref{tab:redundancy}).\\
\begin{mdframed}[backgroundcolor=lightgray!30,%
	linecolor=lightgray!30,%
	innertopmargin = 3pt,%
	innerleftmargin = 3pt,%
	innerrightmargin = 3pt]
For~\ref{rq:guide}, we conclude that feature causes and degrees of blame 
reveal and quantify the impact of features on the desired effect and, this way, 
are able to guide the feature configuration process.
\end{mdframed}

\subsection{Feature Interactions (\ref{rq:interaction})}
Causal reasoning provides a new angle to study feature interactions in configurable systems.
For illustration of how to detect and isolate feature interactions, we perform causal analysis on
the \lrzip subject system, modeling a compression system for which
runtime characteristics of the compression algorithms are of interest.
Since the number of feature causes and their expansions can be both exponential
in the number of features, a direct evaluation of the runtimes and causal analysis
results is difficult. We hence investigate feature interactions through
their degrees of blame as described at the end of Section~\ref{sec:responsibility}.
The subject effect sets $\Effects_{\rho>\thresh_T}$ depend on the runtime 
$\rho\colon \ValidFeat \ra \Real$ in seconds for a configuration compressing a file 
that is obtained as by Siegmund et al.~\cite{SieGreApe15a,KalGreSie2019}
and a runtime threshold $\thresh_T$ (see end of Section~\ref{sec:effects}).
We then focus on $2$-way interactions by investigating potential feature interactions 
between the compression algorithm and compression level responsible to have high runtimes.
For this we compute degrees of feature interaction blame for partial 
configurations $\partial$ where
$\supp(\partial)\in \{\textit{Gzip}, \textit{Lrzip}, \textit{Zpaq}\}\times\{1,\ldots,9\}$ and
$\partial(x)=\true$ for each $x\in\supp(\partial)$.
In the columns of Table~\ref{tab:lrzip}, we show the degrees of feature interaction blames
for thresholds $\thresh_T$ ranging from $\thresh_T=200s$ to $\thresh_T=2\,300s$.
Empty cells correspond to combinations of compression algorithm and level that
do not appear in any cause and thus have zero blame. 
In these cases, we can conclude that no feature interaction takes place.
Higher blame values indicate that the combined responsibility of
the compression algorithm and level has a greater causal impact on runtime. 
Notably, we observe that, with an increasing threshold, the
level of compression is increasingly responsible for longer runtime.
Certain compression algorithms always have runtimes above the
threshold independently of the compression level.
This leads to a configuration blame of zero at any compression level, e.g.,
thresholds~$\thresh_T\leq 600s$ for the \textit{Zpaq} algorithm shown in the upper right
of Table~\ref{tab:lrzip}. Note that, in these cases, \textit{Zpaq} serves as
$1$-way interaction witness according to Definition~\ref{def:fci}.
All greater thresholds for \textit{Lrzip} and \textit{Zpaq} do not have
$1$-way but $2$-way interaction witnesses. That is, being above the runtime threshold
is a result of a feature interaction between the compression algorithm and those compression
levels not showing zero blame.
Notice that the sums of the given feature interaction
blames for $\thresh_T \geq 700s$ that contain algorithms \textit{Lrzip} or \textit{Zpaq}
add up to $1/2$. That is, no other features are to be blamed for exceeding the runtime threshold.

Features have a dedicated meaning and one would hence expect higher runtimes
for higher compression levels. To this end, it seems odd that the feature interaction
of \textit{Lrzip} and compression level 9 is less to blame for higher runtimes than for 
levels 7 and 8.
This indicates an anomaly of the feature interaction between \textit{Lrzip} and the
compression levels 7, 8, and 9. Further investigations on analysis results
and feature causes support these findings: averaged over all measurements of \textit{Lrzip} 
configurations, we observe runtimes of 1\,064.9s at compression level 7 (standard deviation 4.1s), 
1\,181.7s at level 8 (standard deviation 3.2s), and only 830.5s at level 9
(standard deviation 2.6s). Hence, \textit{Lrzip} at level 9 is not causally
relevant for exceeding the execution time threshold of 900s, as the compression level 9 
feature is not contained in any cause with \textit{Lrzip}. However, this insight is
difficult to obtain relying purely on the performance influence model given by $\rho$,
as this would require handcrafted analysis of all 432 analysis results (see Table~\ref{tab:stats}).\\

\begin{mdframed}[backgroundcolor=lightgray!30,%
	linecolor=lightgray!30,%
	innertopmargin = 3pt,%
	innerleftmargin = 3pt,%
	innerrightmargin = 3pt]
For~\ref{rq:interaction} we conclude that feature causes can provide hints for
feature interactions and anomalies arising from them.
Blame measures render themselves promising to quantify the influence of
feature interactions that contribute to certain effects.
\end{mdframed}

\section{Discussion}
In this section, we discuss potential threats to validity of our experiments
and relate our findings to existing work from the literature.

\subsection{Threats to Validity}
A threat to internal validity arises from the correctness of the analysis results
from which we generated the effect sets. While for functional properties this threat
is not crucial due to exact model-checking techniques used in our experiments,
for non-functional properties the results have been partly established using machine learning.
To mitigate this threat, we carefully chose effect-set thresholds such that the effect
sets remain stable also within small threshold variations.
Note that the choice of the effect set has no influence on the applicability of our causality
definitions but only on to what extent causality can serve as an explication.
For blame computations in our experiments, we assumed a uniform distribution over all effects,
taking a developers' perspective where frequencies on how often an
effect occurs in a real-world setting are not yet accessible.
Other distributions could change our quantitative results,
it is unlikely that they would alter our conclusions about causal influences of features
and feature interactions.
To increase the internal validity of our prototype,
we implemented and evaluated several methods to compute causes. These
include a naive brute-force approach and two additional methods to generate prime implicants,
independent from the tool \espresso{}.

Naturally, the choice of subject systems threatens external validity,
which includes the kinds of effect sets on which we evaluate causality.
To alleviate this threat, we included a wide variety of
systems with multiple properties from different areas to our evaluation.
They comprise several real-world software systems often
used to evaluate sampling strategies and performance-modeling approaches.
We further added several community benchmarks from the feature-oriented model-checking 
community as well as a large-scale redundancy system from reliability engineering.

\subsection{Related Work}\label{sec:related}
Various techniques for software defect detection have been proposed in the
literature, ranging from testing~\cite{Mye04a}
and static code analysis~\cite{NieNieHan10a} to model checking~\cite{BK08}.
These techniques have been also extended for analyzing configurable systems
to tackle huge configuration spaces~\cite{Thum14}.
While such methods are able to identify defects and their location,
the challenge of finding root causes for defects remains.
A methodology to identify causes of defects during software development
is provided through \emph{root cause analysis}~\cite{abs1999root,RooHeu04a},
which can be supported by a multitude of techniques for causal reasoning~\cite{Pearl09,Peters2017}.
To the best of our knowledge, the foundations for a combination of configurable
systems analysis and causal reasoning as we presented in this paper have
not yet been addressed in the literature.
In the following, we discuss related work in the fields of configurable
systems analysis and causal reasoning.

\tudparagraph{Configurable systems analysis and explications}
For analyzing configurable software systems, many approaches have been
presented in the last two decades~\cite{Post2008,Thum14}.
There is broad tool support for variability-aware testing and
sampling~\cite{SieRosKuh+12,GYS+18,KalGreSie2019,BeeDamLie19a,KGS+20},
static analysis~\cite{BodTolRib13,Rhein2018,WAS21,VJS+20,VJS+21}, and
model checking~\cite{PlaRya2001,ProveLines13,ClaCorSchobHeyLegRas13,ApeRheWen2013,CDKB18,VanBeeLeg18a}.

There is a substantial corpus of work on determining those features in a configurable
system that are responsible for emerging effects~\citep[e.g.][]{KuhWalGal04a,YilCohPor06a,QuCohRot08a}.
The focus has been mainly on detecting feature interactions~\cite{CalKolMag03a,CalMil06a,ApeAtlBar+2014}.
Siegmund et al.~\cite{SieKolKaeApe2012} and Kolesnikov et al.~\cite{KolesnikovSKGA19} 
describe non-functional feature interactions as
interactions where the composed non-functional property diverges from the aggregation of the individual
contributions of the single features.
Garvin and Cohen~\cite{GarCoh11a} provided a formal definition of feature interaction faults
based on black-box analysis to guide white-box isolation of interaction faults.

An incremental software configuration approach to optimize non-functional properties
has been presented by Nair et al. \cite{NYM+20}, which complements our causality-guided
software configuration we exemplified in Section~\ref{subsec:AVCL} based on
feature causality and precomputed analysis results.

To reduce the size of propositional logic formulas in configurable systems, von Rhein
et al.~\cite{RheGreApe15a} proposed to exclude information about valid configurations
and use two-level logic minimization, e.g., by the \espresso\ heuristics~\cite{McC56a,McGSanBra93a}.
Our DLS method differs from this approach by prioritizing causal information over reduction.

\tudparagraph{Causal reasoning}
Algorithmic reasoning about actual causes following the structural-equation approach
by Halpern and Pearl \cite{HalPearl01-Causes,Halpern2015} is computationally hard in
the general case~\cite{EitLuk02,AlChockHalIv17}.
However, tractable instances such as the Boolean case
have been identified by Eiter and Lukasiewicz~\cite{EitLuk06}.
For deciding whether a partial interpretation is an actual cause in the Boolean
case, Ibrahim and Pretschner presented an approach based on SAT solving~\cite{IbrPre20}.
To compute all causes, their implementation relies on checking causality for all
possible partial interpretations, suffering from an additional exponential
blowup in the number of variables, which we avoid within our approach using prime
implicant computations.

Using test generation methods relying on program trace information,
program locations that are the origin of the defect can be 
identified~\cite{JohBruMel20,RoeFraZel12}.
Analyzing differences between program states of sampled failing and passing executions,
\emph{delta debugging} identifies code positions relevant for an emerging failure~\cite{CleZel05a}.
Similarly, causes for detects can be determined by analyzing counterexample traces~\cite{GroVis03a,BeerBCOT2012}.
Faults can be also located by causal inference on graphs constructed from statement
and test coverage data~\cite{BaaPodHar2010}.

Iqbal et al.\ present a static technique to generate causal models of a given 
configurable system using causal interference
and statistical counterfactual reasoning~\cite{krishna2021cadet}. 
This model is used to detect performance bugs and provide hints for their resolution.
While we focused on actual causality and rigorous analysis,
they are interested in \emph{type causality} to answer more generic questions.

\section{Concluding Remarks}\label{sec:conclusion}
Finding actual causes for an effect event becomes increasingly important in many 
research areas, also driven by high demands from politics and society. 
We introduced a formal definition and algorithms to identify causes 
in configurable systems that relies on counterfactual reasoning
and connections to classical problems of propositional logics and circuit optimization.
We demonstrated their potential by analyzing several subject systems, 
including real-world software systems and popular community benchmarks.
To enable explanations for causes and their impact onto effects,
we proposed explication techniques to concisely represent causes and
quantify the causal impact of features. We showed that our explications
are meaningful and can support the development of configurable software systems
by causality-guided configuration and isolating feature interactions.
With our prototypical implementation, we showed that our algorithms are 
effective on real-world systems of varying sizes and run in reasonable time.

While already shown to be effective, our implementation could be enhanced by directly integrating
feature cause computations into optimized algorithms to compute prime implicants, e.g.,
relying on prime implicant computations at the level of BDDs~\cite{CoudertM92}.
Combined with state-of-the-art BDD libraries such as CUDD~\cite{cudd},
the computation of causes might become feasible for even larger systems than considered in this paper.

Another direction is by enhancing our analysis of feature interaction blames (see Theorem~\ref{thm:fis}) 
with in-depth white-box analyses~\cite{GarCoh11a} to pinpoint root causes in source code 
for a great variety of effect properties using feature causality.

Further applications could be imagined for \emph{context-aware systems} where feature-oriented
formalisms have been shown great applicability \cite{Mauro2016,DKT19,ChrBaiDub20}.
Here, our causal framework could reason about contexts responsible for certain effects,
e.g., in self-adaptive systems~\cite{AssBaiDub21}.
In this vein, \emph{dynamic configurable systems}~\cite{GomHus2003,Dub21} are also
an interesting direction to be considered. In such systems,
features can be activated or deactivated during runtime, e.g., to model
upgrade and downgrade of systems. The detection and isolation of feature interactions 
in dynamic configurable systems is a well-known challenge~\cite{LiuMei09a}.
It is a promising avenue of further work to extend our causal framework
to determine root causes and identify feature interactions in the dynamic setting~\cite{BeerBCOT2012}.\\[-.8em] 
\tudparagraph{Acknowledgments}
  The authors are supported by the \grantsponsor{dfg}{DFG}{http://www.dfg.de}
  through the Collaborative Research Center
  \grantnum[https://perspicuous-computing.science]{dfg}{TRR 248}, the
  \grantnum{exini}{Cluster of Excellence EXC 2050/1} (CeTI, project ID
  390696704, as part of \grantsponsor{exini}{Germany's Excellence Strategy}),
  project \grantnum{dfg}{AP 206/11-1},
  and the Research Training Groups \grantnum{dfg}{QuantLA (GRK 1763)} and
  \grantnum{dfg}{RoSI (GRK 1907)}.
 \clearpage
\bibliographystyle{ACM-Reference-Format}
\bibliography{main}

\end{document}